\documentclass[12pt]{article}

\usepackage[dvips]{color}

\usepackage{epsfig}
 1
\makeatletter
\@addtoreset{equation}{on}
\renewcommand{\theequation}{\thesection.\arabic{equation}}
\makeatother
\begin{document}
\begin{flushright}
HIP-2007-41/TH\\
\end{flushright}
\vspace{0.5cm}
\begin{center}
{\Large \bf
 Meta-stable Vacuum in Spontaneously Broken ${\cal N}=2$
 Supersymmetric Gauge Theory
}
\end{center}
\vspace{8mm}
\begin{center}
\normalsize
{\large \bf  Masato Arai $^a$\footnote{masato.arai@helsinki.fi},
Claus Montonen $^a$\footnote{claus.montonen@helsinki.fi},
Nobuchika Okada $^{b,c}$\footnote{nobuchika.okada@kek.jp}
\vspace{2mm}\\
and
\vspace{2mm}\\
Shin Sasaki $^a$\footnote{shin.sasaki@helsinki.fi}
}
\end{center}
\vskip 1.2em
\begin{center}
{\it
$^a$High Energy Physics Division,
             Department of Physical Sciences,
             University of Helsinki \\
 and Helsinki Institute of Physics,
 P.O.Box 64, FIN-00014, Finland \\
\vskip 1.0em
$^b$Department of Physics, University of Maryland,
College Park,  MD 20742, USA \\
\vskip 1.0em
$^c$Theory Division, KEK, Tsukuba 305-0801, Japan
}
\end{center}
\vskip 0.8cm
\begin{center}
{\large Abstract}
\vskip 0.4cm
\begin{minipage}[t]{16cm}
\baselineskip=19pt
\hskip4mm
We consider an $\mathcal{N}=2$ supersymmetric $SU(2) \times U(1)$
 gauge theory with $N_f=2$ massless flavors and a Fayet-Iliopoulos (FI) term.
In the presence of the FI term, supersymmetry is spontaneously broken
 at tree level (on the Coulomb branch), leaving a pseudo-flat
 direction in the classical potential.
This vacuum degeneracy is removed
 once quantum corrections are taken into account.
Due to the $SU(2)$ gauge dynamics, the effective potential
 exhibits a local minimum at the dyon point,
 where not only supersymmetry but also $U(1)_R$ symmetry is broken,
 while a supersymmetric vacuum would be realized toward infinity
 with the runaway behavior of the potential.
This local minimum is found to be parametrically long-lived.
Interestingly, from a phenomenological point of view,
 in this meta-stable vacuum
 the massive hypermultiplets inherent in the theory
 play the role of the messenger fields
 in the gauge mediation scenario,
 when the Standard Model gauge group is embedded into
 their flavor symmetry.

\end{minipage}
\end{center}
\newpage
%
%
\def\barr{\begin{eqnarray}}
\def\earr{\end{eqnarray}}

\renewcommand{\thefootnote}{\arabic{footnote}}
\setcounter{footnote}{0}
\section{Introduction}
Recently, the possibility that a supersymmetry (SUSY) breaking
vacuum
 is not the global minimum but a local one
 has been proposed by Intriligator, Seiberg and Shih (ISS) \cite{ISS}.
They have investigated an
 ${\cal N} = 1$ SUSY $SU(N_c)$ gauge theory (SUSY QCD).
The number of flavors is taken to be in the range,
 $N_c + 1 \le N_f < {3 \over 2} N_c$,
 so that this theory is described as the infrared free
 magnetic dual theory at low energies and can be analyzed perturbatively.
The effective theory has the same structure as the O'Raifeartaigh
 model, and SUSY is broken with the pseudo-flat directions
 parameterized by meson fields in the dual theory.
The vacuum degeneracy is removed once one-loop corrections
 to the K\"ahler potential are taken into account,
 and a SUSY breaking minimum shows up at the origin
 in the moduli space.
In addition to this minimum, there exist SUSY vacua in this model,
 away from the local minimum.
It has been shown in Ref.~\cite{ISS} that this false vacuum
 can be long-lived and thus meta-stable.

The idea of a meta-stable SUSY breaking vacuum opens up
 a lot of theoretical possibilities for SUSY breaking.
For such a false vacuum, the conventional argument
 using the Witten index is not applicable,
 and it is generally possible for a theory to include
 a local minimum with broken SUSY
 even though the Witten index implies the existence of a SUSY vacuum.
Similarly, the theorem \cite{NeSe} that
 a model with spontaneous SUSY breaking should have an R-symmetry
 is not applicable to a model with a SUSY breaking local minimum.
This feature is welcome from a phenomenological point of view,
 because R-symmetry forbids gauginos to obtain masses.
The R-symmetry should be broken spontaneously or explicitly
 to realize a phenomenologically viable model.
For example, it has been argued \cite{InSeSh2}
 that SUSY breaking in a meta-stable vacuum
 requires only an approximate R-symmetry.
Spontaneous R-symmetry breaking in gauged O'Raifeartaigh
 models \cite{InSeSh2} and
 modified O'Raifeartaigh models \cite{Sh,Fe}
 with meta-stable SUSY breaking vacua have been discussed.

Since the paper by ISS, there have been lots of explorations
 of models with meta-stable vacua.
A meta-stable SUSY breaking vacuum can simplify
 the gauge mediation scenario \cite{GMSB} and several simple models
 have been proposed \cite{AhSe, nomura, DiMa, Al, KaShVo, AbDuJaKh}.
String theory realizations of the meta-stable SUSY breaking vacuum
 have been investigated in Refs.~\cite{OoOo, FrUr, BeGoHeSeSh, Ah,
 EtHaTe, TaWe},
 where a SUSY breaking scale lower than the string scale
 can be realized.

In $\mathcal{N}=1$ SUSY models,
 a meta-stable vacuum can be analyzed only in a weak coupling regime
 in the (effective) theory
 by perturbative means.
Our lack of knowledge about the non-perturbative K\"ahler potential
 prevents us from moving away from the weak coupling limit.
However, in a class of $\mathcal{N}=2$ SUSY gauge theories,
 we can analyze the vacuum structure of a model
 beyond perturbation theory as first demonstrated
 by Seiberg and Witten \cite{s-w1, s-w2},
 using the properties of holomorphy and duality.
In Ref.~\cite{OoOoPa, Pa}, ${\cal N}$= 2 SUSY gauge theories perturbed
 by an appropriate superpotential have been studied
 beyond the perturbative regime.
It has been shown that such perturbed
 $\mathcal{N}=2$ SUSY gauge theories
 can have meta-stable vacua at generic points in the moduli space.

In this paper, we revisit the $\mathcal{N}=2$ SUSY
 gauge theory
 with a Fayet-Iliopoulos (FI) term investigated in Ref.~\cite{arai}.
The model is based on the gauge group $SU(2) \times U(1)$
 with $N_f=2$ massless hypermultiplets.
At the classical level, this theory has SUSY vacua
 on the Higgs branch, at the origin of the Coulomb branch.
Except for near the origin on the Coulomb branch,
 the classical potential possesses SUSY breaking minima
 along a pseudo-flat direction on the Coulomb branch.
 These are far away from the Higgs branch and
 parameterized by moduli parameters, scalars of vector multiplets.
In the quantum theory,
 these pseudo flat directions are removed and a non-trivial local vacuum
 may arise while the SUSY vacua on the Higgs branch
 would remain.

The effective potential along the pseudo flat direction
 can be analyzed beyond perturbation theory
 by using the exact results in $\mathcal{N}=2$ SUSY QCD \cite{s-w1,
 s-w2}.
It is found that the effective potential exhibits
 a local minimum with broken SUSY at the dyon point
 through the $SU(2)$ gauge dynamics and
 also that $U(1)_R$ is dynamically broken there.
The global structure of the potential is determined
 in perturbation theory and the effective potential is found to be of
 the so-called runaway type, namely, the potential energy decreases
 toward infinity where the SUSY vacuum would be realized.
We discuss the vacuum structure of this model in more detail
 and give a rough estimate of the decay rate of the local minimum
 to the runaway vacuum and the SUSY vacua
on the Higgs branch.  We find that this local minimum is
parametrically long-lived
 and thus meta-stable.
Also, we address phenomenological applications of our model.
In fact, in this meta-stable vacuum, the massive hypermultiplets
 inherent in the model play the role of messenger fields
 in the gauge mediation scenario
 when the flavor symmetry among the hypermultiplets
 is gauged as the Standard Model gauge group.

The organization of this paper is as follows.
In \S \ref{vacuum_structure_cl}, the model is defined
 and its classical vacuum structure is studied.
In \S \ref{quantum_theory}, low energy effective Lagrangian
 is derived using the exact results in SUSY QCD.
The effective potential is analyzed in \S \ref{numerical_analysis},
 and we show that the effective potential exhibits a local minimum
 at the dyon singular point due to non-perturbative $SU(2)$ effects.
In \S \ref{lifetime}, we give a rough estimate for the decay rate
 of the local minimum and show the vacuum can be long-lived.
Phenomenological applications of the model are addressed
 in \S \ref{phenomenology}.
The last section is devoted to our conclusion.
Detailed derivations of the effective couplings
 are given in an Appendix.

\section{Vacuum structure of classical theory \label{vacuum_structure_cl}}
We first define our classical Lagrangian and analyze its classical
vacuum\footnote{The complete analysis of the classical
potential for the one flavor case was originally performed in
Ref.~\cite{fayet}.}.
We describe the classical Lagrangian in terms of ${\cal N}=1$ superfields:
 Adjoint chiral superfields $A_i$ and vector superfields $V_i$ in the
vector multiplet
 ($i=1,2$ denote the index of the $U(1)$ and the $SU(2)$ gauge
symmetries, respectively),
 and two chiral superfields $Q^r_I$ and $\tilde{Q}_r^I$
 in the hypermultiplet ($r=1,2$ is the flavor index,
 and $I=1,2$ is the $SU(2)$ color index).
The superfield strength is defined by $W_{i \alpha}
= - \frac{1}{4} \overline{D}^2 (e^{-V_i} D_{\alpha} e^{V_i})$.
The classical Lagrangian is given by
\begin{eqnarray}
{\cal L}&=&{\cal L}_{\rm HM}
           +{\cal L}_{\rm  VM}
           +{\cal L}_{\rm FI} \; ,  \label{eq:lag}
\end{eqnarray}
\begin{eqnarray}
{\cal L}_{\rm HM}
   &=& \int d^4\theta  \left(
       Q_r^\dagger e^{2V_2+2V_1} Q^r
       + \tilde{Q}_r e^{-2V_2-2V_1} \tilde{Q}^{\dagger r}
          \right)  \nonumber \\
   &+& \sqrt{2} \left( \int d^2 \theta
           \tilde{Q}_r \left( A_2+A_1 \right) Q^r + h.c. \right) \; ,
   \label{cl2} \\
{\cal L}_{\rm VM}
        &=&\frac{1}{2\pi} \mbox{Im}\left[ \mbox{tr}
           \left\{\tau_{22}
           \left( \int d^4\theta A_2^\dagger e^{2 V_2} A_2 e^{-2 V_2}
           +\frac{1}{2}\int d^2 \theta  W_2^2
 \right)\right\}\right] \nonumber \\
  &+& \frac{1}{4\pi}{\rm Im}\left[\tau_{11} \left(\int d^4\theta
           A_1^\dagger A_1+\frac{1}{2} \int d^2\theta
           W_1^2 \right)\right] \; , \\
{\cal L}_{\rm FI}
        &=&\int d^4\theta \xi V_1  \; ,      \label{eq:FI}
\end{eqnarray}
where $\tau_{22}=i\frac{4\pi}{g^2}+\frac{\theta}{2\pi}$
 and $\tau_{11}=i\frac{4\pi}{e^2}$ are the gauge couplings
 of the $SU(2)$ and the $U(1)$ gauge interactions, respectively.
Here we use the notation,
 $\mathrm{tr}(T^a T^b)=T(R) \delta^{ab} = \frac{1}{2}\delta^{ab}$
 for the $SU(2)$ generators $T^a$.
The common $U(1)$ charge of the hypermultiplets is normalized to one.
The last term in Eq.~(\ref{eq:lag}) is the FI term
 with a coefficient $\xi$ of mass dimension two.
In what follows, we assume that $\xi>0$.
In general, the FI term also appears in F-term, but the $SU(2)_R$
 symmetry allows us to take a frame so that it appears only in D-term.
Because of this, the $SU(2)_R$ symmetry is explicitly broken down to
 its subgroup $U(1)_{R^\prime}$.
The global symmetry of the theory turns out to be
 $SU(2)_{\rm Left}\times SU(2)_{\rm Right}\times U(1)_{R^\prime}\times
U(1)_R$.
\footnote{Without $U(1)$ gauge symmetry (and the FI term), the flavor
symmetry
 $SU(2)_{\rm Left}\times SU(2)_{\rm Right}$ is enhanced
 to $O(4)$ since the representation of $Q$ and $\tilde{Q}$ are in an
 isomorphic representation of $SU(2)$ gauge group.}

From the above Lagrangian, the classical potential is read off as
\begin{eqnarray}
 V &=&\frac{1}{g^2}\mbox{\rm tr}[A_2, A_2^\dagger]^2
       +\frac{g^2}{2}
 ( q_r^{\dagger} T^a q^r- \tilde{q}_r T^a  \tilde{q}^{\dagger r} )^2
 \nonumber \\
    &+& q_r^\dagger [A_2, A_2^\dagger] q^r
     - \tilde{q}_r [A_2,A_2^\dagger] \tilde{q}^{\dagger r}
    +2 g^2 |\tilde{q}_r T^a q^r |^2  \nonumber \\
   &+&\frac{e^2}{2} \left( \xi+ q_r^\dagger q^r- \tilde{q}_r
       \tilde{q}^{\dagger r} \right)^2
     +2 e^2 |\tilde{q}_r q^r|^2 \nonumber \\
   &+& 2 \left( q_r^\dagger |A_2+A_1|^2 q^r
    +\tilde{q}_r |A_2+A_1|^2 \tilde{q}^{\dagger r} \right) \; ,
\label{classical_potential}
\end{eqnarray}
where $A_2$, $A_1$, $q^r$ and $\tilde{q}_r$ are scalar components
 of the corresponding chiral superfields.

There are supersymmetric vacua in this potential. For example, a solution
\begin{eqnarray}
& A_1 = A_2 = 0\,, \nonumber &\\
& q_r = 0, \ \tilde{q}_1 = (v, 0)\,, \
\tilde{q}_2 = (0,v)\,, \nonumber &\\
& {\xi \over 2} - |v|^2 = 0\,,& \label{SUSY_vacua}
\end{eqnarray}
is a possible supersymmetric vacuum.

\begin{figure}[h]
\begin{center}
\includegraphics[scale=.8]{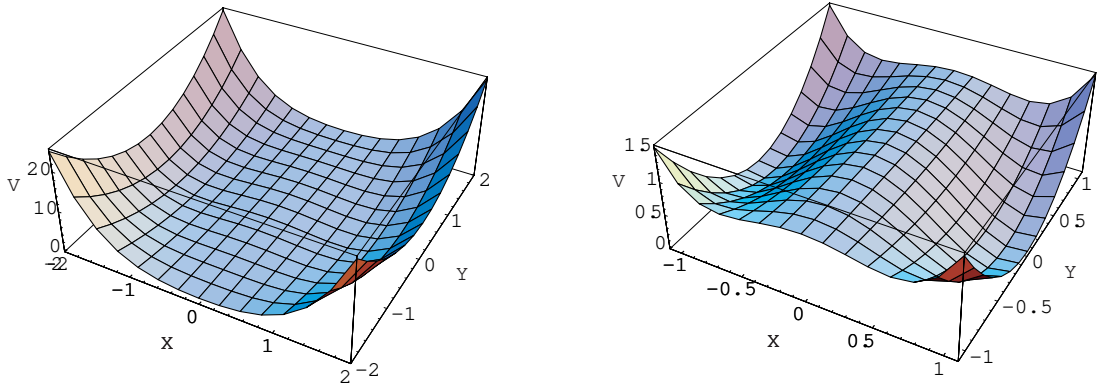}
\end{center}
\caption{Schematic picture of the classical potential. $\xi$ is
taken to be 1 in (\ref{Eq.schematic_classical_potential}).
For $Y>\sqrt{2}\xi$ (left), the minimum is
along $X = 0$.
For $Y<\sqrt{2}\xi$ (right), there are tachyonic
directions along $X$.} \label{schematic_classical_potential}
\end{figure}

Let us then investigate the global structure of the vacuum.
In order to do that,  consider the following
field configuration:
 $q_r = 0$ and $ \tilde{q_1} = (a,0), \  \tilde{q_2} = (0, b), A_2 =
 \mathrm{diag}(a_2/2, - a_2/2)$ and $A_1 = \mathrm{diag}(a_1,a_1)$.
Here, $a_1$ and $a_2$ are complex parameters, and
 $a$ and $b$ are, for simplicity, taken to be real.
Then the potential
(\ref{classical_potential}) is written as
\begin{eqnarray}
V= \frac{g^2}{2} (a^2 -b^2)^2
 + \frac{e^2}{2} (\xi -a^2 - b^2)^2
 + 2 a^2 \left|a_1 + \frac{1}{2} a_2 \right|^2
 + 2 b^2 \left|a_1 - \frac{1}{2} a_2 \right|^2. \label{flat}
\end{eqnarray}
The behavior of the potential (\ref{flat}) is essentially the same as the
 following scalar potential,
\begin{eqnarray}
V = (\xi - X^2)^2  + X^2 Y^2\,, \label{Eq.schematic_classical_potential}
\end{eqnarray}
where $X$ and $Y$ can be regarded as hypermultiplet and vector multiplet
 directions, respectively.
The plot of the potential is depicted in Fig. \ref{schematic_classical_potential}.
The potential has a pseudo flat direction along $Y$ with $X=0$.
The vacua on this flat direction are
 tachyonic along the $X$-direction for $Y<\sqrt{2}\xi$,
 but there are no longer tachyonic directions for $Y>\sqrt{2}\xi$.
In the case (\ref{flat}), a pseudo flat direction is, for instance,
 parameterized by
\begin{eqnarray}
 b=0\,~~~\mbox{\rm and}~~~z=a_1-{1\over 2}a_2\,,  \label{eq:mix1}
\end{eqnarray}
 for $a_1+{1 \over 2}a_2=0$, which corresponds to $X$ and $Y$, respectively.
Along this direction,
 the potential is further minimized with respect to $a$, whose value
 at the minimum is
 given by
\begin{eqnarray}
 a=\left({e^2 \over {1 \over 4}g^2 +
   e^2}\xi\right)^{1 \over 2}\,. \label{eq:mix2}
\end{eqnarray}
In this example, the gauge symmetry $SU(2)\times U(1)$ is broken to
 a linear combination, $U^\prime(1)$.
The potential energy at this minimum is given by
\begin{eqnarray}
 \displaystyle V={\xi^2 \over 2}{e^2g^2 \over 4e^2+g^2}\,.
\end{eqnarray}
Therefore, the supersymmetry is broken at the minimum along the pseudo flat
 direction by the non-zero FI parameter.

We are now interested in what happens to this pseudo flat direction in quantum theory.
Since $a_1$ and $a_2$ have $U(1)_R$ charge $+2$,
 the $U(1)_R$ symmetry is broken by these
 non-zero vacuum expectation values (VEVs).
We expect that the pseudo-flat direction is lifted up,
 once quantum corrections are taken into account,
 and some non-degenerate vacua would appear
 after the effective potential is analyzed.
This naive expectation seems natural, if we notice that the above
 potential energy is described by the bare gauge couplings,
 which should be replaced by the effective ones
 (non-trivial functions of the moduli parameters) in the effective theory.
In the following sections, we will show that quantum corrections
 actually remove the vacuum degeneracy
 and leave two vacua on the Coulomb branch , one of which is a local minimum
breaking both SUSY and $U(1)_R$ symmetry and the other is a runaway
vacuum.

In addition to this runaway SUSY vacuum in the Coulomb
branch, there would be other SUSY vacua. It is known that
there are no quantum corrections
 on the Higgs branch \cite{ArPlSe},
 and we thus expect that the classical SUSY vacua
 (\ref{SUSY_vacua}) survive after quantum corrections
 are taken into account.
At these SUSY vacua, the hypermultiplets have
 very small VEVs because of the theoretical consistency condition
 $\xi \ll \Lambda^2$ (see the next section).
On the Coulomb branch, except for very near the origin,
 we have a pseudo flat direction as shown above
 and no tachyonic direction towards the Higgs branch.
For this reason, we will analyse the quantum theory
 along the Coulomb branch and will see how the effective
 potential can be modified along the pseudo flat directions.

\section{Quantum theory \label{quantum_theory}}
In this section, we describe the low energy Wilsonian effective
 Lagrangian of our theory.
The detailed derivation of the effective action is found in
Ref.~\cite{arai}.
Here, we briefly summarize the results for the convenience of the reader.

In order to derive the exact low energy effective action
 ${\cal L}_{\rm EXACT}$, which is described by the light fields, the
dynamical scale
 and the coefficient of the FI term $\xi$,
 we need to integrate the action to zero momentum.
However, this is a highly non-trivial task.
Without the FI term, the theory has ${\mathcal N}=2$ SUSY, which can be
 utilized to integrate out massive degrees of freedom.
In our model, this is not the case
 because the SUSY is broken at the classical level.
In the following discussion, suppose that the coefficient $\xi$,
 the order parameter of SUSY breaking,
 is much smaller than the dynamical scale $\Lambda$ of the $SU(2)$ gauge
interaction.
Then we can consider the effective action up to the leading order in $\xi$.
The exact effective Lagrangian, if it could be obtained,
 can be expanded in the parameter $\xi$ as
\begin{eqnarray}
{\cal L}_{\rm EXACT}
        ={\cal L}_{\rm SUSY}
         +\xi {\cal L}_1 + {\cal O}(\xi^2)\,.
         \label{eq:eff}
\end{eqnarray}
Here, the first term ${\cal L}_{\rm SUSY}$
 is the exact effective Lagrangian containing full SUSY quantum
corrections. The second term is the leading term in $\xi$. Since
$\xi$ is a constant and has mass dimension 2, ${\cal L}_1$ should
 be a gauge-invariant quantity having mass dimension 2.
This simple consideration tells us that the second term is nothing but
 the FI term.
 \footnote{The exactness of the FI term is also discussed by using the
 harmonic superspace formalism in Ref. \cite{AK}.}
Analyzing the effective Lagrangian up to the leading order in $\xi$,
 we obtain the effective potential to order $\xi^2$.
The coefficient of $\xi^2$ in the effective potential includes
 the full SUSY quantum corrections.
Therefore, to achieve our aim, what we need to analyze the effective
potential
 is nothing but the effective Lagrangian ${\cal L}_{\rm SUSY}$.

Except for the FI term, the classical $SU(2)\times U(1)$ gauge theory
 has a moduli space, which is parameterized by $a_2$ and $a_1$.
On this moduli space except at the origin, the gauge symmetry is
broken to
 $U(1)_c\times U(1)$.
\footnote{In this paper, we study the Coulomb
 branch, not the mixed branch like in (\ref{eq:mix1}) and (\ref{eq:mix2}), on which it is
 difficult to analyze the effective potential.
However, at low energies, the $SU(2)$ gauge coupling is much larger
 than the $U(1)$ gauge coupling, so that
 the solution (\ref{eq:mix1}) and (\ref{eq:mix2})
 is approximately that of the Coulomb branch.
}
Here $U(1)_c$ denotes the gauge symmetry in the Coulomb phase
 originating from the $SU(2)$ gauge symmetry.
Before discussing the effective action of this theory,
 we should make clear how to treat the $U(1)$ gauge interaction part.
In the following analysis, this part is, as usual, discussed
 as a cut-off theory.
\footnote{
 There is a possibility that
 a non-trivial fixed point and a strong coupling phase
 exist in QED \cite{miransky}.
 This problem is non-trivial, and is outside our scope
 (see also Ref.~\cite{AK} for related discussions).}
Thus, the Landau pole $\Lambda_L$ is inevitably introduced
 in our effective theory, and the defining region
 of the modulus parameter $a_1$ is constrained
 to lie within the region $|a_1|< \Lambda_L$.
Because of this constraint, the defining region for the modulus parameter
 $a_2$ is found to be also constrained to be in the same region,
 since the two moduli parameters are related to each other
 through the hypermultiplets.
We take the scale $\Lambda_L$ to be much larger than
 the dynamical scale of the $SU(2)$ gauge interaction $\Lambda$,
 so that the $U(1)$ gauge interaction is always weak
 in the defining region of moduli space.
Note that, in our framework, we implicitly assume that
 the $U(1)$ gauge interaction has no effect on the $SU(2)$ gauge dynamics.
This assumption will be justified in the following discussion concerning
 the monodromy transformation (see Eq.~(\ref{eq:mono})).

We first discuss the general formulae for the effective Lagrangian
 ${\cal L}_{\rm SUSY}$, which
 consists of two parts described by light vector multiplets
 and hypermultiplets,
 ${\cal L}_{\rm SUSY}={\cal L}_{\rm VM}+{\cal L}_{\rm HM}$.
At low energies,
 the ${\cal N}=2$ effective Lagrangian of the vector multiplet part,
 ${\cal L}_{\rm VM}$, includes
 the superfield $A_2$ of the unbroken Abelian subgroup of $SU(2)$ and
 the Abelian superfield $A_1$.
The effective action consistent with ${\cal N}=2$ SUSY and all the
symmetries
 in our theory is given by
\begin{eqnarray}
{\cal L}_{\rm VM}
        =\frac{1}{4\pi}
           \mbox{\rm Im}\left\{
           \sum_{i,j=1}^{2}\left(\int d^4\theta
           \frac{\partial F}{\partial A_i} A_i^\dagger
         + \int d^2\theta \frac{1}{2}
           \tau_{ij} W_i W_j \right) \right\} ,
\end{eqnarray}
where $F(A_2,A_1,\Lambda,\Lambda_L)$ is the prepotential, which is
 a function of moduli parameters
 $a_2$, $a_1$, the dynamical scale $\Lambda$,
 and the Landau pole $\Lambda_L$.
Note that the effective coupling $\tau_{12}(=\tau_{21})$ appears through
 the quantum corrections.
The effective coupling $\tau_{ij}$ is defined as
\begin{eqnarray}
\tau_{ij}=
   \frac{\partial^2 F}{\partial a_i \partial a_j} \; \; (i,j=1,2) .
   \label{eq:coupling}
\end{eqnarray}
The part ${\cal L}_{\rm HM}$ is
 described by a light hypermultiplet
 with appropriate quantum numbers $(n_e,n_m)_n$,
 where $n_e$ is the electric charge, $n_m$ is the magnetic charge,
 and $n$ is the $U(1)$ charge.
This part should be added to the effective Lagrangian
 around a singular point in moduli space,
 since the hypermultiplet is expected to be light there
 and enjoys correct degrees of freedom in the effective theory.
The explicit description is given by
\begin{eqnarray}
{\cal L}_{\rm HM}
 &=& \int d^4 \theta \left(
      M_r^{\dagger} e^{2 n_m V_{2D}+2 n_e V_2 + 2 n V_1}M^r
      +\tilde{M}_r e^{-2 n_m V_{2D}-2 n_eV_2-2 n V_1} \tilde{M}^{r\dagger}
              \right)  \nonumber \\
 &+& \sqrt{2} \left( \int d^2\theta \tilde{M}_r (n_m A_{2D}+n_e A_2+n
A_1) M^r
               +h.c. \right) \; ,
\label{superpotential}
\end{eqnarray}
where $M^r$ and $\tilde{M}_r$ denote light quark, light monopole
 or light dyon hypermultiplet,
 that is, the light BPS states,
 and $V_{2D}$ is the dual gauge field of $U(1)_c$.
Since the $U(1)$ gauge coupling is weak and does not affect the $SU(2)$
 gauge dynamics, the flavor symmetry is effectively that of
 ${\cal N}=2$ SUSY QCD.
Recalling that $a_1$ plays a role of the hypermultiplet mass if it has
 vacuum expectation value, for vanishing VEV of $a_1$,
 the light BPS states belong to a spinor representation
 of $SO(4)\sim SU(2)_-\times SU(2)_+$ \cite{jackiw,s-w2}.
A non-zero vacuum expectation value of
 $a_1$ breaks the symmetry down to $SU(2)_-\times U(1)_+$.
At the quantum level, the global $U(1)_R$ symmetry is anomalous and
 the resultant anomaly-free symmetry turns out to be
 ${\bf Z}_8\subset U(1)_R$ \cite{s-w2}.

In order to obtain an explicit description of the effective Lagrangian,
 let us consider the monodromy transformations of our theory.
Suppose that the moduli space is parameterized by the vector multiplet
scalars
 $a_2$, $a_1$ and their duals $a_{2D}$, $a_{1D}$
 which are defined as $a_{iD}=\partial F/ \partial a_i$ ($i=1,2$).
These variables are transformed into their linear combinations
 by the monodromy transformation.
In our case, the monodromy transformations form a subgroup
 of $Sp(4,\mbox{\bf R})$, which leaves the effective Lagrangian invariant,
 and the general formula is found to be \cite{marino3}
\begin{eqnarray}
 \left(\begin{array}{c}
       a_{2D} \\  a_{2}  \\  a_{1D} \\  a_1  \end{array}  \right)
 \rightarrow
 \left(\begin{array}{c}
       \alpha a_{2D}+\beta a_2+p a_1   \\
       \gamma a_{2D}+\delta a_2+q a_1  \\
       a_{1D}+p(\gamma a_{2D}+\delta a_2)
        -q(\alpha a_{2D}+\beta a_2)-pqa_1\\ a_1
\end{array}  \right)\,,
  \label{eq:mono}
\end{eqnarray}
where
$
\left(\begin{array}{cc}
\alpha & \beta  \\ \gamma & \delta \end{array} \right)
 \in SL(2,\mbox{\bf Z})
$
 and $p,q \in\mbox{\bf Q}$.
Note that this monodromy transformation for the combination
 $(a_{2D},a_{2},a_1)$ is exactly the same as that
 for SUSY QCD with massive quark hypermultiplets,
 if we regard $a_1$ as the common mass of the hypermultiplets
 such that $m= \sqrt{2} a_1$.
This fact means that the $U(1)$ gauge interaction part only plays
 the role of the mass term for the $SU(2)$ gauge dynamics.
This observation is consistent with our assumptions.
On the other hand, the $SU(2)$ dynamics plays an important role
 for the $U(1)$ gauge interaction part,
 as can be seen from the transformation law of $a_{1D}$.
This monodromy transformation is also used to derive dual variables
 associated with the BPS states.
As a result, the prepotential of our theory turns out to be
 essentially the same as the result in Ref.~\cite{s-w2}
 with the additional relation $m=\sqrt{2}A_1$,
\begin{eqnarray}
F(A_2,A_1,\Lambda, \Lambda_L)
   =F_{SU(2)}^{(SW)}(A_2, m,\Lambda)
    \Bigg{|}_{m=\sqrt{2}A_1}+C A_1^2\,,
    \label{eq:pre}
\end{eqnarray}
where the first term on the right hand side
 is the prepotential of ${\cal N}=2$ SUSY QCD
 with hypermultiplets having the same mass $m$,
 and $C$ is a free parameter.
The freedom of the parameter $C$ is used to determine
 the scale of the Landau pole
 relative to the scale of the $SU(2)$ dynamics.

The effective potential can be obtained from the action, after
 eliminating auxiliary fields
\footnote{
 We presuppose that the potential is described by the proper variables
 associated with the light BPS states.
 For instance, the variable $a_2$ is understood implicitly as $-a_{2D}$,
 when we consider the effective potential for the monopole.}
%
\begin{eqnarray}
V &=& \frac{b_{22}}{2\det b}\xi^2
      +S(a_2,a_1)\left\{(|M^r|^2 - |\tilde{M}_r|^2)^2
                 +4|M^r \tilde{M}_r|^2\right\} \nonumber \\
 &+&  2 T(a_2,a_1)(|M^r|^2 + |\tilde{M}_r|^2)
      -U(a_2,a_1) (|M^r|^2-|\tilde{M}_r|^2)\,, \label{pot-q}
\end{eqnarray}
where
 $|M^r|^2=M^rM_r^{\dagger},~|\tilde{M}_r|^2=\tilde{M_r}\tilde{M}^{r\dagger}$,
$b_{ij}=(1/4 \pi) \mbox{Im} \tau_{ij}$ is the effective coupling and
 $\det b\equiv b_{22} b_{11} -b_{12}^2$.
The functions $S$, $T$ and $U$ are defined as
\begin{eqnarray}
S(a_2,a_1)&=&\frac{1}{2b_{22}}
            +\frac{(b_{12}-nb_{22})^2}
                  {2b_{22}\det b}\,, \\
T(a_2,a_1)&=&|a_2+na_1|^2, \\
U(a_2,a_1)&=&\frac{b_{12}-nb_{22}}{\det b}\xi\,.
\end{eqnarray}
Solving the stationary conditions with respect to the hypermultiplets,
 we have the following three solutions:
\begin{eqnarray}
1. \;  & & M = \tilde{M} =0;  \; \;
  V=\frac{b_{22}}{2\det b}\xi^2\,,  \label{eq:sol1} \\
2. \; & & |M^r|^2=-\frac{2T-U}{2S}, \; \tilde{M}=0; \; \;
  V=\frac{b_{22}}{2\det b}\xi^2-S |M^r|^4\,,
       \label{eq:sol2} \\
3. \; & & M=0, \; |\tilde{M}_r|^2 = -\frac{2T+U}{2S};  \; \;
  V=\frac{b_{22}}{2\det b}\xi^2 -S |\tilde{M}_r|^4\,.  \label{eq:sol3}
\end{eqnarray}
The solution Eq.~(\ref{eq:sol2}) or Eq.~(\ref{eq:sol3}),
 in which the light hypermultiplet acquires
 a vacuum expectation value,
 is energetically favored, because $\det b >0$ and $S(a_2,a_1)>0$.
Since the hypermultiplet appears in the theory as the light BPS state
 around the singular point in moduli space,
 a potential minimum is expected to emerge there.
In addition, the points (\ref{eq:sol2}) and  (\ref{eq:sol3})
 are stable in the $M, \tilde{M}$ directions.
This is because they are unique solutions and have lower energy
 than the point (\ref{eq:sol1}),
 and the potential at infinity in $M, \tilde{M}$
 space is dominated by the $M^4, \tilde{M}^4$ terms.
On the other hand, the solution Eq.~(\ref{eq:sol1}) describes
 the potential energy away from the singular points,
 which smoothly connects with the solution
 Eq.~(\ref{eq:sol2}) or Eq.~(\ref{eq:sol3}).

It was shown that the effective potential is described by the periods
 $a_{2D}$, $a_{2}$ and the effective gauge coupling $b_{ij}$.
The periods are the same as that of massive SUSY QCD.
Although there are some different descriptions of the periods
 it is convenient for our purpose
 to write them as integral representations \cite{marino3} and
 to write the effective coupling $\tau_{ij}$
 in terms of the Weierstrass functions.

We first review how to obtain the periods $a_{2D}$ and  $a_2$.
The elliptic curve of ${\cal N}=2$ SUSY QCD with two hypermultiplets
 having the same mass $m$ was found to be \cite{s-w2}
\begin{eqnarray}
y^2=x^2(x-u)-\frac{\Lambda^4}{64}(x-u)
       +\frac{\Lambda^2}{4}
       m^2x-\frac{\Lambda^4}{32}m^2\,,
 \label{eq:curve}
\end{eqnarray}
where $u={\rm Tr}A_2^2$ is identified with the modulus parameter.
In this case, the mass formula of the BPS state
 with the quantum numbers $(n_e, n_m)_n$
 is given by $M_{\rm BPS}=\sqrt{2}| n_m a_{2D}+n_e a_2+n m/\sqrt{2}|$.
If $\lambda$ is a meromorphic differential on the
 curve Eq.~(\ref{eq:curve}) such that
\begin{eqnarray}
\frac{\partial \lambda}{\partial u}
       =\frac{\sqrt{2}}{8\pi}\frac{dx}{y}\,,
\end{eqnarray}
the periods are given by the contour integrals
\begin{eqnarray}
  a_{2D}=\oint_{\alpha_1}\lambda\,, \;
  a_{2}=\oint_{\alpha_2}\lambda\,,
 \label{eq:period}
\end{eqnarray}
where the cycles $\alpha_1$ and $\alpha_2$ are defined
 so as to encircle $e_2$ and $e_3$, and $e_1$ and $e_3$, respectively,
 which will be given explicitly later on (see eq.(\ref{eq:root})).
The meromorphic differential is given by
\begin{eqnarray}
\lambda_{SW}
     &=&-\frac{\sqrt{2}}{4\pi}
        \frac{ydx}{x^2-\frac{\Lambda^4}{64}}
      =-\frac{\sqrt{2}}{4\pi}\frac{dx}{y}
        \left[ x-u  +\frac{m^2\Lambda^2}
         {4\left(x+\frac{\Lambda^2}{8} \right)}\right]\,.
     \label{eq:f2}
\end{eqnarray}
The differential has a single pole at
 $x=-\frac{\Lambda^2}{8}$ and
 the residue is given by
\begin{eqnarray}
\mbox{\rm Res}\lambda_{SW}
   =\frac{1}{2\pi i}(-1)\frac{m}{\sqrt{2}}\,.
\end{eqnarray}

We calculate the periods by using the Weierstrass normal form
 for later convenience.
In this form, the algebraic curve is rewritten
 in new variables $x=4X+\frac{u}{3}$ and $y=4Y$, such that
\begin{eqnarray}
Y^2 = 4X^3-g_2X-g_3
   &=&4(X-e_1)(X-e_2)(X-e_3)\,,\\ \nonumber
\sum_{i=1}^3e_i&=&0\,,
\end{eqnarray}
where $g_2$ and $g_3$ are explicitly written by
\begin{eqnarray}
g_2&=&\frac{1}{16}\left(\frac{4}{3}u^2
                   +\frac{\Lambda^4}{16}-m^2
                    \Lambda^2\right)\,,\\
g_3&=&\frac{1}{16}\left(
                   \frac{m^2 \Lambda^4}{32}
                   -\frac{u}{12}m^2\Lambda^2
                   -\frac{u\Lambda^4}{96}
                   +\frac{2u^3}{27}\right)\,.
\end{eqnarray}
Converting the Seiberg-Witten differential,
 Eq. (\ref{eq:f2}), into the Weierstrass normal form
 and substituting it into Eq.(\ref{eq:period}),
 we obtain the integral representations of the periods as follows
 ($a_{2D}$ and $a_{2}$ are denoted by $a_{21}$ and $a_{22}$, respectively):
\begin{eqnarray}
 a_{2i}=-\frac{\sqrt{2}}{4\pi}
    \left(-\frac{4}{3}uI_1^{(i)}+8I_2^{(i)}
    +\frac{m^2\Lambda^2}{8}
    I_3^{(i)}
    \left(c \right)\right)\,,
\end{eqnarray}
where $c$ is the pole of the differential,
 given by $c=-\frac{u}{12}-\frac{\Lambda^2}{32}$.
The integrals $I_1^{(i)},I_2^{(i)}$ and $I_3^{(i)}$ are defined as
\begin{eqnarray}
I_1^{(i)}=\frac{1}{2}\oint_{\alpha_i}\frac{dX}{Y}\,, \; \;
I_2^{(i)}=\frac{1}{2}\oint_{\alpha_i}\frac{XdX}{Y}\,, \; \;
I_3^{(i)}(c)=\frac{1}{2}
            \oint_{\alpha_i}\frac{dX}{Y(X-c)}\,.
\end{eqnarray}
The roots $e_i$ of the polynomial defining the cubic
 are chosen so as to lead to the correct asymptotic behavior
 for large $|u|$,
\begin{eqnarray}
a_{2D}(u)
 \sim i\frac{2}{2\pi}\sqrt{2u} \log\frac{u}{\Lambda^2}\,, \; \; \;
a_2(u)\sim\frac{\sqrt{2u}}{2}\,,
\end{eqnarray}
A correct choice is the following:
\begin{eqnarray}
e_1 &=&\frac{u}{24}-\frac{\Lambda^2}{64}
     -\frac{1}{8}\sqrt{u+\frac{\Lambda^2}{8}
                  +\Lambda m}
                 \sqrt{u+\frac{\Lambda^2}{8}
                  -\Lambda m}\,,
                 \nonumber \\
e_2 &=&\frac{u}{24}-\frac{\Lambda^2}{64}
     +\frac{1}{8}\sqrt{u+\frac{\Lambda^2}{8} + \Lambda m}
                 \sqrt{u+\frac{\Lambda^2}{8} - \Lambda m}\,,
                 \label{eq:root}\\
e_3 &=&-\frac{u}{12}+\frac{\Lambda^2}{32}\,.
                 \nonumber
\end{eqnarray}
Fixing the contours of the cycles relative to the positions
 of the poles, which is equivalent to fixing the $U(1)$ charges
 for the BPS states, the final formulae are given by
\begin{eqnarray}
a_{2i}=-\frac{\sqrt{2}}{4\pi}
    \left(-\frac{4}{3}uI_1^{(i)}+8I_2^{(i)}
    +\frac{m^2\Lambda^2}{8}
    I_3^{(i)}
    \left(-\frac{u}{12}
    -\frac{\Lambda^2}{32}\right)\right)
    -\frac{m}{\sqrt{2}}\delta_{i2}\,,
    \label{eq:period2}
\end{eqnarray}
with the integrals $I_s^{(1)} \; (s=1,2,3)$ explicitly given by
\begin{eqnarray}
I_1^{(1)}&=&\int_{e_2}^{e_3}\frac{dX}{Y}
          = \frac{iK(k^\prime)}{\sqrt{e_2-e_1}}\,,
            \label{eq:formula1} \\
I_2^{(1)}&=&\int_{e_2}^{e_3}\frac{XdX}{Y}
          = \frac{ie_1}{\sqrt{e_2-e_1}}K(k^\prime)
            +i\sqrt{e_2-e_1}E(k^\prime)\,,
            \label{eq:formula2} \\
I_3^{(1)}&=&\int_{e_2}^{e_3}\frac{dX}{Y(X-c)}
             = \frac{-i}{(e_2-e_1)^{3/2}}
            \left\{
            \frac{1}{k+\tilde{c}}K(k^\prime)
            +\frac{4k}{1+k}
            \frac{1}{\tilde{c}^2-k^{2}}
            \Pi_1\left(\nu,\frac{1-k}{1+k}
                \right)
            \right\}\,,
            \label{eq:formula3}
\end{eqnarray}
where $k^2 = \frac{e_3-e_1}{e_2-e_1}$,
      $k^{\prime 2}=1-k^2=\frac{e_2-e_3}{e_2-e_1}$,
      $\tilde{c}= \frac{c-e_1}{e_2-e_1}$,
and $\nu=-\left(\frac{k+\tilde{c}}{k-\tilde{c}}\right)^2
          \left(\frac{1-k}{1+k}\right)^2$.
The formulae for $I_s^{(2)}$ are obtained from $I_s^{(1)}$
 by exchanging the roots $e_1$ and $e_2$.
In Eqs.~(\ref{eq:formula1})-(\ref{eq:formula3}),
 $K$, $E$, and $\Pi_1$ are the complete elliptic integrals \cite{higher}
 given by
\begin{eqnarray}
K(k)&=&\int_0^1 \frac{dx}
     {\left[(1-x^2)(1-k^2x^2)\right]^{1/2}}\,,
       \\ \nonumber
E(k)&=&\int_0^1 dx\left(\frac{1-k^2x^2}{1-x^2}
     \right)^{1/2}\,,
       \\ \nonumber
\Pi_1(\nu,k)&=&\int_0^1\frac{dx}
     {[(1-x^2)(1-k^2x^2)]^{1/2}(1+\nu x^2)}\,.
\end{eqnarray}

Next let us consider the effective coupling defined in
 Eq.~(\ref{eq:coupling}).
A detailed derivation of the effective couplings is given in the Appendix.
The effective couplings $\tau_{22}$ and $\tau_{12}$
 are obtained by
\begin{eqnarray}
\tau_{22}&=&\frac{\partial a_{2D}}{\partial
a_2}=\frac{\omega_1}{\omega_2}\,,\\
\tau_{12}&=&
\frac{\partial a_{2D}}{\partial a_1}
           \Bigg{|}_u
          -\tau_{22}
          \frac{\partial a_{2}}{\partial a_1}
           \Bigg{|}_u
=-\frac{2z_0}{\omega_2}\,,
         \label{eq:tau12}
\end{eqnarray}
where $\omega_i$ is the period of the Abelian differential,
\begin{eqnarray}
\omega_i=\oint_{\alpha_i}\frac{dX}{Y}=2I_1^{(i)} \; ~(i=1,2)\,,
\end{eqnarray}
and $z_0$ is defined as
\begin{eqnarray}
z_0=-\frac{1}{\sqrt{e_2-e_1}}F(\phi,k); \; \;
 \sin^2\phi=\frac{e_2-e_1}{c-e_1}\,.
\end{eqnarray}
Here $F(\phi,k)$ is the incomplete elliptic integral of
 the first kind given in (\ref{incomp}).

The effective coupling $\tau_{11}$ is described
 in terms of the Weierstrass function.
First consider the period $a_{1D}$
 by using the Riemann bilinear relation \cite{riemann},
\begin{eqnarray}
\oint_{\alpha_1}\phi\oint_{\alpha_2}\omega
     -\oint_{\alpha_1}\omega\oint_{\alpha_2}\phi
     =2\pi i\sum_{n=1}^{N_p}
      \mbox{\rm Res}_{x_n^+}\phi
      \int_{x_n^-}^{x_n^+}\omega\,, \label{eq:bilinear}
\end{eqnarray}
where $\phi$ and $\omega$ are meromorphic and
 holomorphic differentials, respectively,
 $N_p$ is the number of poles ($N_p=1$ in our case),
 and $x_n^{\pm}$ are poles of $\phi$ on the positive and
 negative Riemann sheets.
Substituting
 $ \phi={\partial \lambda_{SW}}/{\partial a_1}$
 and $ \omega={\partial \lambda_{SW}}/{\partial a_2}$
 into Eq.~(\ref{eq:bilinear}), we obtain
\begin{eqnarray}
a_{1D}=-\sum_{n=1}^{N_p} \int_{x_n^-}^{x_n^+}
     \lambda_{SW} + \tilde{C} \; , \label{eq:mass}
\end{eqnarray}
where $\tilde{C}$ is a constant independent of $a_2$.
The effective coupling $\tau_{11}$ is obtained
 by differentiating Eq.~(\ref{eq:mass}) with respect to $a_1$
 with $a_2$ fixed.
The integral in Eq. (\ref{eq:mass}) after the differentiation
 can be evaluated by the uniformization method
 discussed in the Appendix.
After regularizing the integral by using the freedom
 of the constant $\tilde{C}$,
 we finally obtain (see also the Appendix for details)
\begin{eqnarray}
\tau_{11}=-\frac{1}{\pi i}\left[\log\sigma(2z_0)
           +\frac{4z_0^2}{\omega_2}
           I_2^{(1)}\right] + C ,
         \label{eq:tau11}
\end{eqnarray}
where $\sigma$ is the Weierstrass sigma function,
 and $C$ is the constant in Eq.~(\ref{eq:pre}).

\begin{figure}[h]
\begin{center}
\leavevmode
  \epsfysize=4.7cm
  \epsfbox{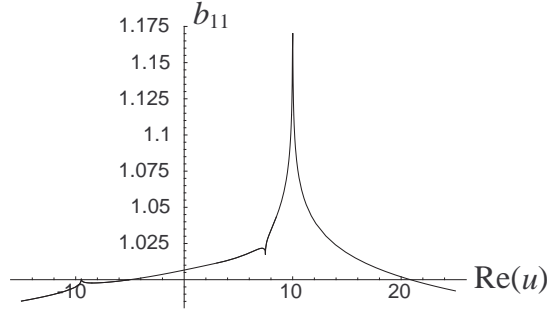} \\
\caption{The effective gauge coupling $b_{11}$ for $a_1=3/\sqrt{2}$
 along the real $u$ axis.}
\label{coupling}
\end{center}
\end{figure}

We now define the Landau pole associated with the $U(1)$ interaction.
In the ultraviolet region far away from the origin of the moduli space,
 the effective coupling is dominated by the $U(1)$ gauge interaction
 since the $SU(2)$ interaction is asymptotic free and small.
As we expect, the gauge coupling $b_{11}$ is found to be
 a monotonically decreasing function of the large $|a_1|$
 with fixed $u$, and vice versa
 (see, for example, Fig.~\ref{coupling} in the case of fixed $a_1$).
The Landau pole is defined as $|a_1|=\Lambda_L$
 at which $b_{11}=0$.
The large $\Lambda_L$ required in our assumption
 is realized by taking an appropriate value for $C$.
In the following analysis, we fix $C =4 \pi i $,
 which corresponds to $\Lambda_L = 10^{17-18}$
 in units of $\Lambda$.

%
%
\begin{figure}
\begin{center}
\leavevmode
  \epsfysize=4.7cm
  \epsfbox{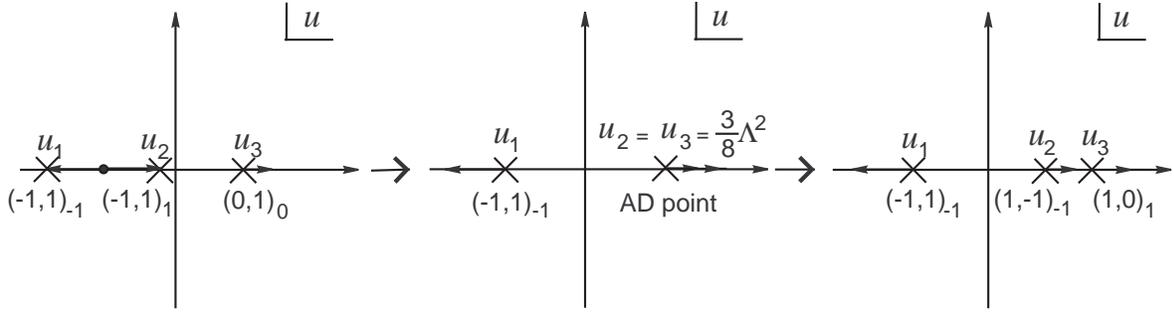} \\
\caption{Flow of the singular points as ${\rm Re}(a_1)$ increases with
 ${\rm Im}(a_1)=0$.}
\label{evo2}
\end{center}
\end{figure}

\section{Numerical analysis of the effective potential
\label{numerical_analysis}}
In this section, we examine the effective potential minimum numerically.
As explained in the previous section, the minimum is expected to appear
 at the singular point since it is energetically favored
 due to the non-zero condensation (see Eq.~(\ref{pot-q}))
 of the light BPS state such as a quark, monopole or dyon
 with appropriate quantum number $(n_e,n_m)_n$.
Thus, let us first investigate the singular points,
 and then analyze the effective potential at the singular point.

\begin{figure}
\begin{center}
\leavevmode
  \epsfysize=4.2cm
  \epsfbox{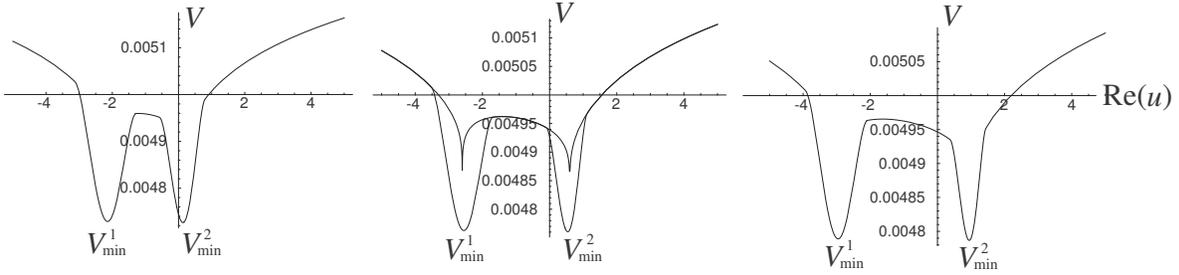} \\
\caption{The effective potential for $a_1=0.3$(left), $a_1=0.4$(middle)
 and $a_1=0.5$(right) on the real $u$ axis.}
\label{plot}
\end{center}
\end{figure}
The singular points on the moduli space is determined
 by the cubic polynomial \cite{s-w2}.
The solutions of the cubic polynomial give the positions of the singular
points
 in the $u$-plane.
In the $N_f=2$ case with the same hypermultiplet masses, the solution is
 easily obtained as
\begin{eqnarray}
u_1=-m\Lambda-\frac{\Lambda^2}{8}{\Bigg |}_{m=\sqrt{2}a_1}\,, \;
u_2=m\Lambda-\frac{\Lambda^2}{8}{\Bigg |}_{m=\sqrt{2}a_1}\,, \;
u_3=m^2+\frac{\Lambda^2}{8}{\Bigg |}_{m=\sqrt{2}a_1}\,. \label{descriminat}
\end{eqnarray}
The flow of the singular points with respect to
 the real hypermultiplet mass is sketched in Fig.~\ref{evo2}.
For $a_1=0$, the singular points appear
 at $u_1=u_2=-\Lambda^2/8$ and $u_3=\Lambda^2/8$.
Here, at $u=-\Lambda^2/8$, two singular points coincide.
For non-zero $a_1>0$,
\footnote{For ${\rm Im}(a_1) =0$, it is enough to consider only the case
$a_1 > 0$,
 since the result for $a_1 < 0$ can be obtained
 by exchanging $u_1 \leftrightarrow u_2$,
 as can be seen from the first two equations in Eq.~(\ref{descriminat}).}
 this singular point splits into two singular points $u_1$ and $u_2$,
 which correspond to the BPS states with quantum numbers
 $(-1,1)_{-1}$ and $(-1,1)_1$, respectively.
As $a_1$ is increasing, these singular points, $u_1$ and $u_2$,
 are moving to the left and the right on the real $u$-axis, respectively.
The two singular points, $u_2$ and $u_3$, collide and coincide
 at the so-called Argyres-Douglas (AD) point \cite{argyres}
 ($u=\frac{3\Lambda^2}{8}$) for $a_1=\frac{\Lambda}{2\sqrt{2}}$,
 where it is believed that the theory becomes superconformal.
As $a_1$ increases further, there appear two singular points
 $u_2$ and $u_3$ again,
 and the quantum numbers of the corresponding BPS states,
 $(-1,1)_1$ at $u_2$ and $(0,1)_0$ at $u_3$, change into $(1,-1)_{-1}$
and $(1,0)_1$,
 respectively.
The singular point $u_3$ is then moving away to the right faster
 than $u_2$.\\

\begin{figure}
\begin{center}
  \epsfysize=5.2cm
  \epsfbox{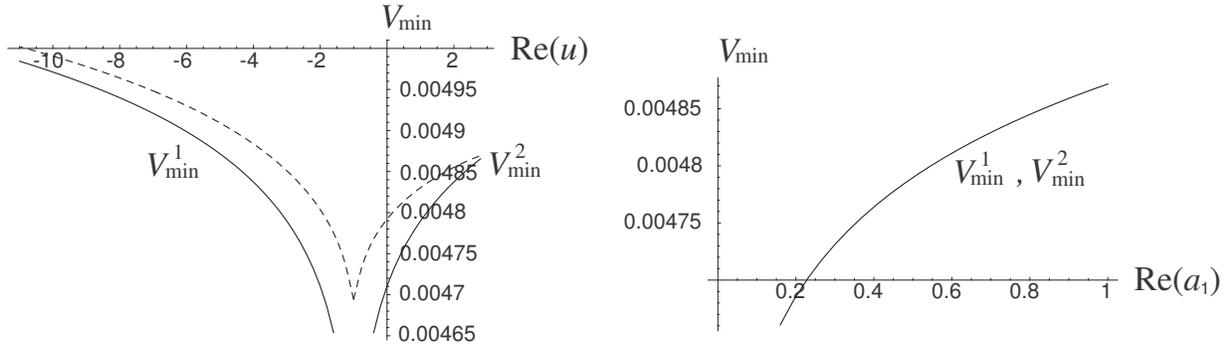} \\
\caption{The evolution of the potential minima $V_{\rm min}^1$ and
 $V_{\rm min}^2$ at the singular points $u_1$ and $u_2$, respectively, as
 $a_1$ varies on the real $u$-axis (left) and on the real $a_1$ axis
 (right). The solid (dashed) curve shows the plots with(without) dyon
condensation.}
\label{graph6}
\end{center}
\end{figure}

Now let us examine the effective potential at the singular point.
First note that the effective potential is a function of $u$ and $a_1$,
 $V(a_2(u,a_1),a_1)$ (see (\ref{eq:sol2}) and (\ref{eq:sol3})).
Furthermore, (\ref{descriminat}) tells us that
 the singular point is completely determined by the value of $a_1$,
 and therefore the potential at the singular point
 is a function of $a_1$ only.
In the following, we investigate the effective potential
 at some fixed value of $a_1$, and see how the minimum appears
 at the singular point.
Then we examine the evolution of the minimum by varying $a_1$.
In our numerical analysis, we take $\Lambda=2\sqrt{2}$ and $\xi=0.1$.

\noindent \underline{(i) $0 \le \mathrm{Re} (a_1) < \frac{\Lambda}{2
\sqrt{2}},
\quad \mathrm{Im} (a_1) = 0
$}
\\
\\
The effective potentials for several values of $a_1$ in the range,
 $0<a_1<\Lambda/2\sqrt{2}$ (corresponding to the left figure in
 Fig. \ref{evo2}), are depicted in Fig. \ref{plot}.
The potential minima, $V_{\rm min}^1$ and $V_{\rm min}^2$,
 appear at two singular points $u_1$ and $u_2$, respectively, while
 there is no minimum at the singular point $u_3$ since the monopole
 condensation is too small for the potential to have a minimum.
In the middle figure the top and the bottom curves show the effective
 potential without and with the dyon condensations, respectively.
The cusps are smoothed out in the bottom curve, which means that
 the correct degrees of freedom in the theory are considered.
The two minima in Fig. \ref{plot} approach each other and their values
 become smaller as $a_1$ decreases.
Such a behavior can be shown in Figs. \ref{graph6}.
The left figure shows the evolution of the potential minima,
 $V_{\rm min}^1$ and $V_{\rm min}^2$ at the dyon
 singular points as $a_1$ changes on the real $u$ axis.
The top and the bottom curves are plots without and with dyon
 condensates, respectively.
From this figure, one sees that the condensation lowers the potential
energy.
The right figure shows that the evolution of the potential minima along
 the real $a_1$ axis.
In this plot, the behaviors at the two dyon singular points completely
 coincide because of the symmetry, $a_1 \rightarrow -a_1$ (see
(\ref{descriminat})).
From the analysis we find that
 the potential is bounded from below, at least along the real $u$ axis,
 and it is expected that
 there is a (local) minimum at $u\rightarrow -\Lambda^2/8$ and
$a_1\rightarrow 0$.

\begin{figure}[t]
\begin{center}
\leavevmode
  \epsfysize=4cm
  \epsfbox{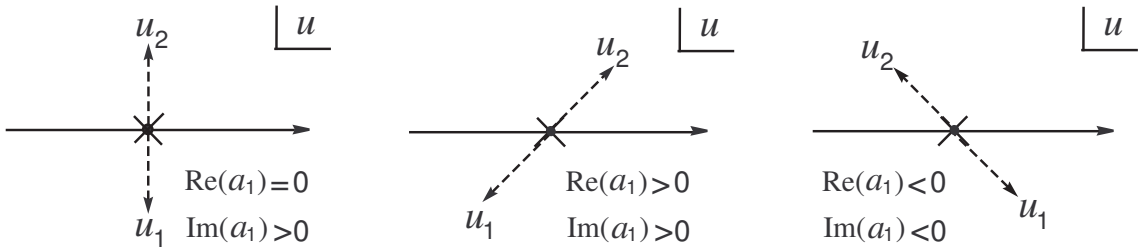} \\
\caption{Flow of the singular points $u_1$ and $u_2$ for general values
of $a_1$.}
\label{evo2-im}
\end{center}
\end{figure}
\begin{figure}[t]
\hspace{-1cm}
  \epsfysize=5.5cm
 \hspace{2cm}
  \epsfbox{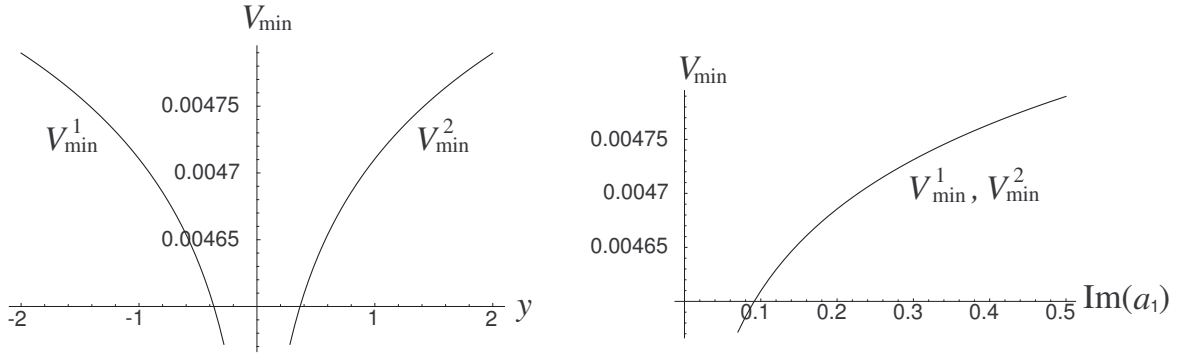} \\
\caption{The evolution of the minima $V_{\rm min}^1$ and $V_{\rm min}^2$
 at $u_1$ and $u_2$, respectively, with varying the pure imaginary
 part of $a_1$ along $u=-1+iy$ axis(left) and the imaginary $a_1$ axis.}
\label{graph-im2}
\end{figure}
\begin{figure}
\begin{center}
\begin{tabular}{cc}
\leavevmode
 \epsfysize=5.8cm
\epsfbox{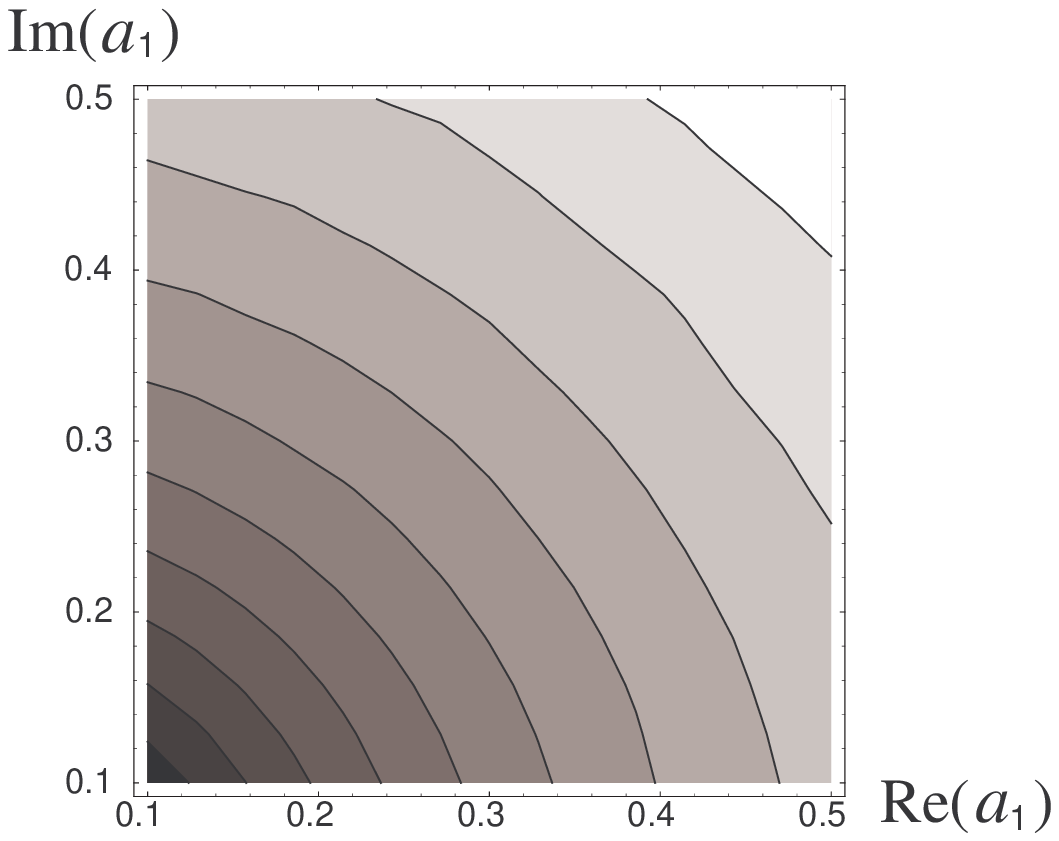} &
\leavevmode
 \epsfysize=6cm
\epsfbox{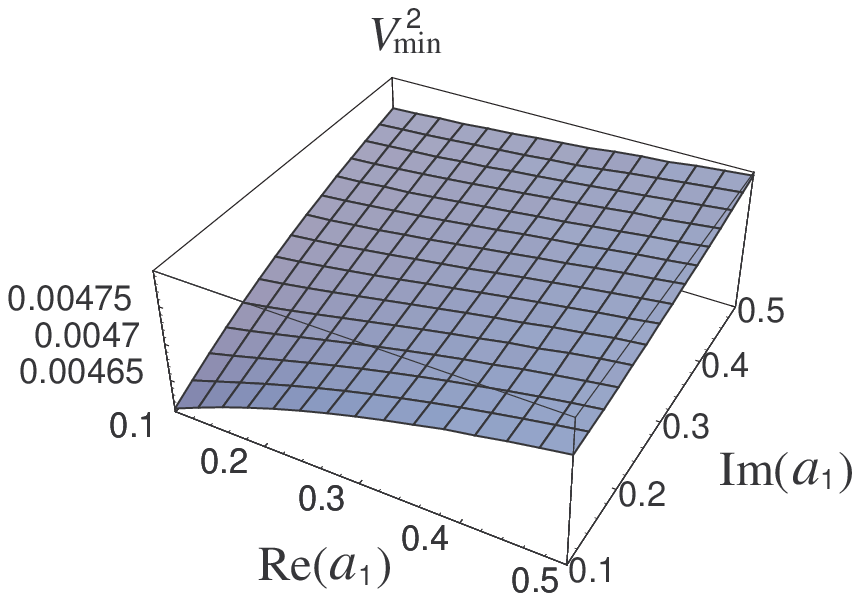}\\
\end{tabular}
\caption{
The contour and the 3D plots of the effective potential at the singular
 point $u_2$ as a function of $a_1$.}
\label{contour}
\end{center}
\end{figure}

\noindent \underline{(ii) $\mathrm{Re} (a_1) = 0, \quad \mathrm{Im}
(a_1) \not= 0$}
\\
\\
Next we examine the effective potential
 for a complex value of $a_1$ around $a_1=0$.
For our purpose, it is sufficient to investigate small values of
 ${\rm Im}(a_1)$ near ${\rm Re}(a_1)=0$
 since we want to know whether the effective potential
 is bounded from below or not at the point ${\rm Re}(a_1)=0$.
Once again let us go back to the flow of the singular points.
 Fig. \ref{evo2-im} shows the flow of the singular points
 $u_1$ and $u_2$ for several complex values of $a_1$.
The left figure shows the flow as ${\rm Im}(a_1)$
 increases.
The singular points $u_1$ and $u_2$ are moving in opposite directions
 along $u=-\Lambda^2/8=-1$ axis.
The middle(right) figure shows the flow of the singularities for
 the ${\rm Re}(a_1)>0, {\rm Im}(a_1)>0 $ (${\rm Re}(a_1)<0, {\rm
Im}(a_1)<0$) case.
The plots of the potential corresponding to these flows are shown
 in Fig. \ref{graph-im2} and \ref{contour}.
The left figure in Fig. \ref{graph-im2}, corresponding to the left
 figure in Fig. \ref{evo2-im} shows the evolution of the potential
 minima for two dyons along the $u=-1$ axis.
The right figure is the same plot, but along the ${\rm Im}(a_1)$ axis.
Note that in the latter plot, the evolution of the two dyon minima
 completely coincide as in the case of the right figure in Fig.
\ref{graph6}.
These two dyon points roll down to the point $a_1\rightarrow
 0$ ($u\rightarrow -1$), and the potential near
 $a_1=0$ is bounded from below in the pure imaginary direction of $a_1$.
%
%
Fig. \ref{contour} shows the contour and 3D plots of the effective
potential at
 the $u_2$-dyon point as a function of ${\rm Re}(a_1)>0$ and ${\rm
Im}(a_1)>0$.
The dark(light) color shows lower(higher) value of the effective
 potential.
The effective potential is invariant under
 ${\rm Re}(a_1)\rightarrow -{\rm Re}(a_1)$ and/or
 ${\rm Im}(a_1)\rightarrow -{\rm Im}(a_1)$,
 and the plot for other parameter range of $a_1$
 is obtained through this invariance.
The plot of the effective potential at $u_1$-dyon point is obtained
 by exchanging $a_1\rightarrow -a_1$.
In conclusion, the point $u\rightarrow -1$ and $a_1\rightarrow 0$
 is expected to be a local vacuum.

However, note that our description is not applicable  for very
 small $|a_1|$,
 since the condensations of the two dyon states are going to
 overlap with each other (see Fig.~\ref{plot}).
Unfortunately, we have no knowledge about the correct description
 of the effective theory in this situation.
Nevertheless, we conclude that there must appear a local minimum
 with broken SUSY in the limit $a_1 \rightarrow 0$:
In this limit, the effective potential without the dyon condensations
 is depicted in the left figure in Fig.~\ref{graph6}.
We find that a potential minimum appears at $u=-\Lambda^2/8=-1$,
 and the value of the effective potential at the cusp
 is non-zero, $V \simeq 0.0047 >0 $.
If we had the correct description of the effective theory for $a_1=0$,
 this cusp might be smoothed out.
However, there is no reason for SUSY to be restored at $u=-1$,
 because the correct effective theory must have no singularity
 in the K\"ahler metric.
Therefore, there is the promising possibility
 of the appearance of a local minimum with broken SUSY
 at $u=-1$ and $a_1=0$.
Note that in this local minimum the global ${\bf Z}_8\subset U(1)_R$
 symmetry is broken down to ${\bf Z}_4$.

\begin{figure}[t]
\hspace{-32.5cm}
\begin{tabular}{ccc}
  \epsfysize=7cm
  \epsfbox{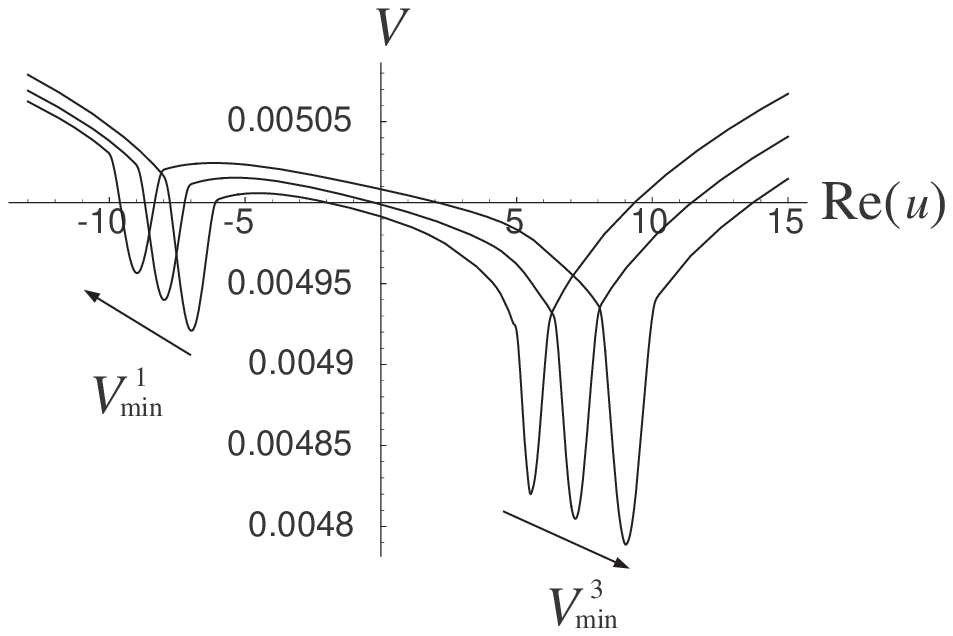} &&
  \epsfysize=4.7cm
  \epsfbox{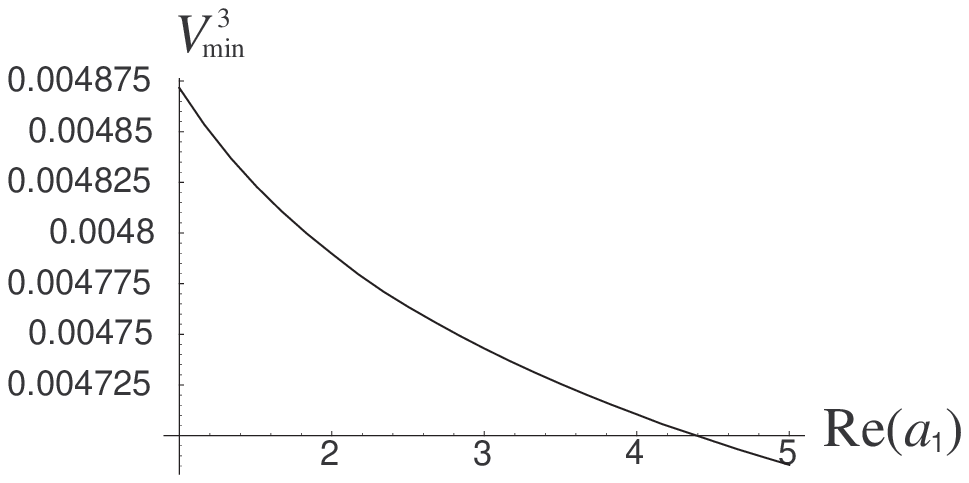}
\end{tabular}
\caption{The evolution of the potential minima at the singular points as
 $a_1$ varies on the real $u$-axis (left) and on the real $a_1$ axis
(right).}
\label{graph11}
\end{figure}
%
\begin{figure}[t]
\vspace{-4cm}
\hspace{-30cm}
  \epsfysize=9.2cm
  \epsfbox{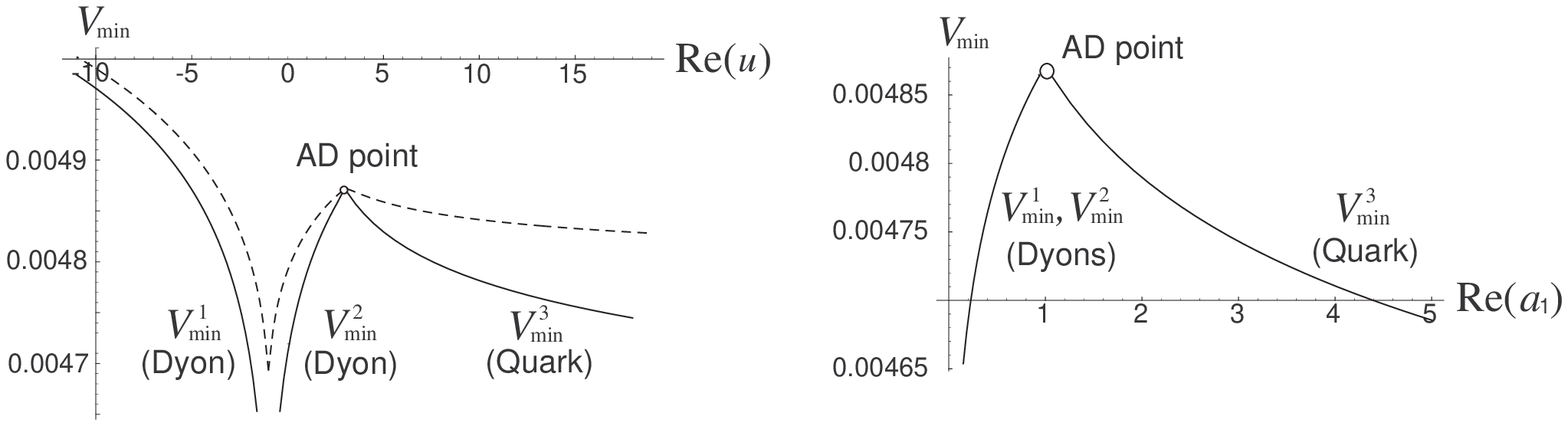} \\
\caption{The effective potential energy at each singular point.}
\label{graph10}
\end{figure}

\noindent \underline{(iii) $\mathrm{Re} (a_1) > \frac{\Lambda}{2 \sqrt{2}}$}
\\
\\
Let us get back to the case of $\mbox{Im}(a_1)=0$.
For $a_1 >\frac{\Lambda}{2 \sqrt{2}}$, the effective potential has two
minima,
 $V_{\rm min}^1$ and $V_{\rm min}^3$ at two singular points $u_1$ and
$u_3$.
The dyon condensation is too small for the effective potential
 to have a minimum at $u_2$.
The plot of the effective potential is shown in Fig.~\ref{graph11}.
While the evolution of the potential energy with the singular point $u_1$
 is the same as for $0 < a_1 < \frac{\Lambda}{2 \sqrt{2}}$,
 the potential energy on the quark singular point at $u_3$
 is monotonically decreasing, as $a_1$ is increasing.
Thus, there is a runaway direction
 along the flow of the quark singular point.
We can find the same global structure along the flow of the quark
singular point
 for general complex $a_1$ values.

The evolutions of the potential energies
 according to the flows of the singular points along the real $u$-axis
 are simultaneously plotted in Fig.~\ref{graph10}.
The global structure of the effective potential
 is of the runaway type.
However, we found the promising possibility that
 there exists a local minimum with broken SUSY in the theory.
Precisely speaking, since there is no well-defined vacuum
 in the runaway direction, this minimum with broken SUSY is
 the unique and promising candidate for the vacuum in the theory.
Unfortunately, we have no knowledge of the correct description
 about the effective theory around the degenerate dyon point.

Finally, we would like to comment on the possible SUSY vacua
 which are present on the Higgs branch at the classical level.
As we have mentioned in \S \ref{vacuum_structure_cl},
 SUSY vacua at the point $a_1 = a_2 = 0$ would exist even at the quantum level.
At the classical level, the SUSY breaking vacua on the pseudo flat
 direction roll down to the SUSY vacua on the Higgs branch
 near the origin $|a_1|, |a_2| \sim \sqrt{\xi}$.
In our analysis of the Coulomb branch, the effective
 potential (\ref{pot-q}) does not give such a picture.
Instead, we have found that the effective potential
 realizes a local minimum at $u\sim \Lambda^2 \gg \xi$,
 such that the minimum is far away from the Higgs branch.
In order to realize the SUSY vacua on the Higgs branch,
 which would exist even in the quantum level,
 appropriate hypermultiplets should be introduced
 in the effective action,
 while we, unfortunately, do not know
 the correct treatment of such degrees of freedom.

\section{Lifetime of the local minimum \label{lifetime}}
In this model, we have found two possible kinds of SUSY vacua:
A runaway vacuum at infinity in moduli
 space on the Coulomb branch and the vacua on the Higgs branch.
Here we will estimate the decay rate from the local vacuum to these SUSY vacua.

First we estimate it for the runaway vacuum. As we discussed
earlier, consistency requires that we restrict the moduli space to a
region bounded by the Landau pole. When the boundary of moduli space
is located far away from the dynamical scale, the potential energy
at the boundary is almost zero, and the true, almost SUSY, vacuum of
the theory lies somewhere in this region. If our world is trapped in
the local minimum we found in the previous sections, it will
eventually decay to this approximately supersymmetric vacuum. The
decay rate is expected to be very small, as the potential barrier is
very wide.

As analyzed in the previous section, the effective potential
 can be described as a function of the modulus parameter $a_1$.
In our calculation, the effective potential is treated
 in the triangle approximation \cite{DuJe}.
Let us take the path in the direction of ${\rm Re}(a_1)$:
 climbing up from the local minimum ($a_1=0$)
 to the AD point ($a_1=\Lambda/2\sqrt{2}$),
 then rolling down to Landau pole point ($a_1=\Lambda_L$).
This is similar to the situation in the ISS model.

\begin{figure}[htb]
\begin{center}
\includegraphics[scale=.55]{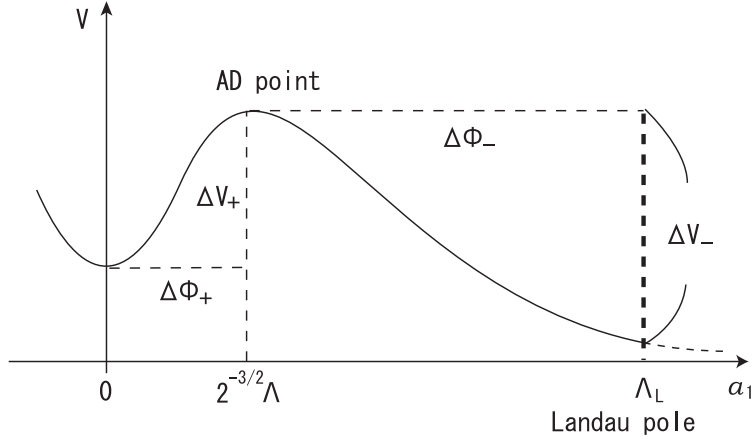}
\end{center}
\caption{Schematic picture of the effective potential}
\label{potential}
\end{figure}

In the triangle approximation, parameters characterizing the potential are
\begin{eqnarray}
\Delta V_{\pm}, \quad \Delta \Phi_{\pm},
\end{eqnarray}
where $\Delta V_{\pm}$ and $\Delta \Phi_{\pm}$
 are the difference of potential height
 and the distances between local/Landau pole points
 and potential barrier (see Fig.~\ref{potential}).
Following reference \cite{DuJe}, we define
\begin{eqnarray}
 \lambda_{\pm} \equiv \frac{\Delta V_{\pm}}{\Delta \Phi_{\pm}}, \quad
  c \equiv \frac{\lambda_{-}}{\lambda_{+}} = \frac{\Delta
 V_{-}}{\Delta V_{+}} \frac{\Delta \Phi_{+}}{\Delta \Phi_{-}}.
\end{eqnarray}
In our case,
\begin{eqnarray}
\Delta \Phi_{+} \sim \Lambda, \ \Delta \Phi_{-} \sim \Lambda_{L},
\end{eqnarray}
and the height of the effective potential is controlled by
 the SUSY breaking order parameter $\xi$ as
\begin{eqnarray}
  V \sim \xi^2 .
\end{eqnarray}
Through numerical analysis, the ratio $\Delta V_{-}/ \Delta V_{+}$
 is estimated to be $\mathcal{O}(10)$,
 so that the condition of Eq.~(13) in Ref.~\cite{DuJe} can be satisfied,
\begin{eqnarray}
 \left( \frac{\Delta V_{-}}{\Delta V_{+}} \right)^{\frac{1}{2}}
 \ge \frac{2 \Delta \Phi_{-}}{\Delta \Phi_{-} - \Delta \Phi_{+}} =
\frac{2}{1 - \frac{\Delta \Phi_{+}}{\Delta \Phi_{-}}} \sim 2.
\label{triangle_condition}
\end{eqnarray}
Here we have used
\begin{eqnarray}
\frac{\Delta \Phi_{+}}{\Delta \Phi_{-}}
\sim \frac{\Lambda}{\Lambda_L} \ll 1\,.
\end{eqnarray}
Since for our choice of parameters, $ \Lambda_L =10^{17-18}$ in
units of $\Lambda$, we can safely use the formula of the bounce
action \cite{DuJe},
\begin{eqnarray}
B = \frac{32 \pi^2}{3} \frac{1 + c}{(\sqrt{1 + c} - 1)^4}
 \frac{\Delta \Phi_{+}^4}{\Delta V_{+}}\,.
\end{eqnarray}
Because the parameter $c$ is very small
\begin{eqnarray}
 c = \frac{\lambda_{-}}{\lambda_{+}}
 = \frac{\Delta V_{-}}{\Delta V_{+}}
 \frac{\Delta \Phi_{+}}{\Delta \Phi_{-}}
 \sim 10 \frac{\Lambda}{\Lambda_L} \ll 1,
\end{eqnarray}
we find
\begin{eqnarray}
B \sim  \frac{\Lambda_L^4}{\xi^2}
 \gg \frac{\Lambda^4}{\xi^2}  \gg 1.
\end{eqnarray}
Here we used the condition $\Lambda^2 \gg \xi$
 for our analysis in the previous section
 to be theoretically consistent.
As a result, the decay rate per unit volume $\Gamma/V \sim e^{-B}$
 from the local minimum to the Landau pole point
 is very small, and the vacuum at the local minimum
 is very long-lived, i.e. meta-stable, as expected.
Although the final formula seems to indicate
 that the decay rate becomes zero in the limit
 $\Lambda_L \rightarrow \infty$,
 it is, in fact, non-zero due to the barrier penetration
 from the local minimum to the runaway direction.

Next we estimate the decay rate from the local minimum
 to the SUSY vacua present on the Higgs branch.
The most conservative path to such SUSY vacua from the local one
 is first climbing up to the origin $a_1 = a_2 = 0, q = \tilde{q} = 0$
 and then rolling down to the SUSY vacuum $ a_1 = a_2 = 0, q, \tilde{q}
\sim \sqrt{\xi}$.
In this situation, the potential parameters are estimated
from the numerical analysis to be
\begin{eqnarray}
& & \Delta \Phi_{+} = \Lambda = 2 \sqrt{2}, \ \Delta \Phi_{-} =
\sqrt{\xi} = \sqrt{0.1}, \nonumber \\
& & \Delta V_{+} = 0.00891, \ \Delta V_{-} = 0.00421.
\label{numerical_potential}
\end{eqnarray}
These parameters do not satisfy the condition
(\ref{triangle_condition}), so that we can not use the formula (13)
 in Ref. \cite{DuJe}.
Instead, we can use another formula (20) in Ref. \cite{DuJe},
\begin{eqnarray}
B = \frac{\pi^2 \lambda_{+}^2 R_T^3}{96}  \left[
- \beta_{+}^3 + 3 c \beta_{+}^2 \beta_{-} + 3 c \beta_{+} \beta_{-}^2
- c^2 \beta_{-}^3 \right]\,.
\end{eqnarray}
Here
\begin{eqnarray}
\beta_{\pm} \equiv \sqrt{\frac{8 \Delta \Phi_{\pm}}{\lambda_{\pm}}}\,,
~~~R_T = \frac{1}{2} \left(
\frac{\beta_{+}^2 + c \beta_{-}^2}{c \beta_{-} - \beta_{+}}
\right)\,.
\end{eqnarray}
The bounce action is estimated to be
 $B \sim \mathcal{O} (10^6) \gg 1$ from the numerical values of
the potential parameters (\ref{numerical_potential}).
Thus the decay rate from the local minimum to
the SUSY vacua can be very small.

\section{Application to phenomenology \label{phenomenology}}
We have found a meta-stable vacuum with broken SUSY
 and also broken $U(1)_R$ symmetry in the previous sections.
Here we address the application of our model to phenomenology.
Supersymmetric extensions of the Standard Model have been considered
 as one of the most promising ways to solve the gauge hierarchy
 problem in the Standard Model.
Since any supersymmetric partners of the Standard Model
 particles have not been observed yet,
 supersymmetry should be broken at low energies.
The origin of supersymmetry breaking and its mediation to
 the supersymmetric version of the Standard Model
 are still prime questions in particle physics.
As mentioned several times before,
 in order to obtain a realistic model,
 $U(1)_R$ symmetry breaking is necessary as well as
 breaking of supersymmetry.
This is because $U(1)_R$ symmetry forbids gauginos
 to obtain masses.
From this point of view, the meta-stable vacuum we have found
 is suitable for phenomenology.
Furthermore, the model automatically provides
 the structure necessary in the gauge mediation scenario.

First, let us give a brief review on the gauge mediation
 scenario \cite{GMSB}.
The basic structure of this scenario is described as
 the messenger sector superpotential,
\begin{eqnarray}
  W= S \tilde{\Phi} \Phi,
  \label{messenger}
\end{eqnarray}
 where $S$ is a gauge singlet chiral superfield,
 and $\tilde{\Phi}$ and $\Phi$ are a vector-like pair of
 chiral superfields, so-called messenger fields,
 which are charged under the Standard Model gauge group.
Suppose that both the scalar component and the F-component of
 the singlet superfield $S$ develop VEVs
 so that SUSY and also $U(1)_R$ symmetry are broken.
Through quantum corrections with messenger fields,
 gauginos and scalar partners of the Standard Models particles
 obtain soft SUSY breaking masses,
\begin{eqnarray}
  M_{\rm soft} \sim \frac{\alpha_{\rm SM}}{4 \pi}
  \frac{\langle F_S \rangle }{\langle S \rangle},
\end{eqnarray}
where $\langle S \rangle$ and $\langle F_S \rangle$
 are VEVs of the scalar and the F-component of the superfield $S$,
 $\alpha_{\rm SM}$ stands for the Standard Model gauge coupling.

Now we return to our model.
At the meta-stable vacuum, $u\rightarrow -\Lambda^2/8$ and $a_1
\rightarrow 0$,
 the model possesses flavor symmetry $SU(2)_{-} \times SU(2)_{+}$
\cite{s-w2}. The BPS states, dyon hypermultiplet, which describe the
low energy effective theory
 around the
 meta-stable vacuum, belong to a doublet under $SU(2)_{+}$.
Note that the model also includes the other massless hypermultiplets
 ($\tilde{D}$ and $D$) doublets under $SU(2)_{-}$ at the singular
 point, $u=\Lambda^2/8$.
However, they can no longer be massless at the other singular point
and
 they exist as massive states in the meta-stable vacuum.
One can understand such a structure in the classical theory. In the
classical superpotential in Eq. (\ref{cl2}),
\begin{eqnarray}
 W=\sqrt{2}\tilde{Q}_{Ir} (A_2+A_1)^I_{~J} Q^{Jr}\,,
\end{eqnarray}
considering that the Cartan part of
 $A_2 \sim {\rm diag}(A_2/2,-A_2/2)$
 is left in the low energy effective theory,
 there are two moduli points, $a_2/2+a_1=0$ and $-a_2/2+a_1=0$, where
 the hypermultiplets are massless.
But at one of the moduli points,
 some hypermultiplets are massless, but the others are massive and are
 integrated out.
The same structure should be realized at the quantum level.
Now the massive hypermultiplet would have the following form of the
superpotential,
\begin{eqnarray}
 W = \sqrt{2} \tilde{D}_r (n_m A_D+n_e A+n A_1) D^r,
 \label{messenger2}
\end{eqnarray}
with certain quantum numbers $(n_e,n_m)_n$ as in
Eq.~(\ref{superpotential}).
Since $\langle A_D \rangle \sim \Lambda$
 at the local minimum we have found,
 the massive hypermultiplets are heavy and integrated out
 from the low energy effective theory.
However, once we take supersymmetry breaking effects into account
 at the minimum, this superpotential is found to play
 an important role in phenomenology.

Supersymmetry is broken at the local minimum and so
 the F-component of $A_D$ develops a VEV characterized
 as $\langle F_{A_D} \rangle \sim \xi$, so that Eq.~(\ref{messenger2})
 has the same structure as Eq.~(\ref{messenger})
 with $A_D= \langle A_D \rangle + \theta^2 \langle F_{A_D} \rangle
       \sim \Lambda + \theta^2 \xi$.
Therefore, when the $SU(2)$ flavor symmetry is weakly gauged as,
 for example, the $SU(2)$ weak gauge group in the Standard Model,
 the hypermultiplets, $\tilde{D}$ and $D$, play the role
 of messenger fields and the $SU(2)$ gaugino and all doublet scalars
 in the supersymmetric Standard Model obtain masses
 through the gauge mediation such as
\begin{eqnarray}
  M_{\rm soft} \sim \frac{\alpha_2}{4 \pi}
  \frac{\xi}{\Lambda} ,
\end{eqnarray}
where $\alpha_2$ is the $SU(2)$ weak gauge coupling.
Suitable choices of model parameters,
 supersymmetry breaking order parameter $\xi$ and
 the messenger scale $\Lambda$,
 lead to phenomenologically favored values for
 soft supersymmetry breaking masses around 1 TeV.

In order to obtain a more realistic phenomenological model,
 it is necessary to extend our model
 so as to provide a larger flavor symmetry.
For example,
 an ideal choice would be an  $SU(5)$ flavor symmetry,
 whose subgroup can be gauged as the Standard Model gauge group
 $SU(3)_C \times SU(2)_L \times U(1)_Y \subset SU(5)$
 so that all gauginos and scalar partners obtain masses.
To implement such a large global symmetry into
 a supersymmetric gauge theory,
 a model should be based on a general $SU(N)$ ($ N > 2$) gauge group
 with an appropriate number of flavors.
It is non-trivial to construct such a model more suitable
 for phenomenology, and we leave this issue for future works.

Our SUSY breaking model is based on $\mathcal{N}=2$ SUSY gauge
 theories, while the supersymmetric Standard Model is a chiral
 theory and should obey $\mathcal{N}=1$ supersymmetry.
It may be somewhat unusual to realize such a setup in four dimensions.
As a natural realization, we can consider
 a $\mathcal{N}=1$ five-dimensional brane world scenario,
 where the SUSY breaking sector resides in the bulk
 while the SUSY Standard Model sector resides on a ``3-brane''.
The Lagrangian for the bulk fields is described
 in terms of $\mathcal{N}=2$ SUSY theory in four-dimensional point of view,
 while that for the brane fields obeys only $\mathcal{N}=1$ supersymmetry.
If we could extend our four-dimensional model to a five-dimensional
one, such a natural phenomenological model would be realized.

\section{Conclusion \label{discussion}}
We have investigated an $\mathcal{N}= 2$ supersymmetric gauge theory
 based on the gauge group $SU(2) \times U(1)$
 with $N_f=2$ flavors and the FI term associated
 with the $U(1)$ gauge group.
Thanks to the exact results in $\mathcal{N}=2$ supersymmetric
 gauge theories, we can analyze the model beyond perturbation theory with
 respect to the $SU(2)$ gauge coupling,
 but as a perturbation with respect to the FI term
 smaller than the $SU(2)$ dynamical scale.
We have found that the effective potential exhibits
 a local SUSY breaking minimum at the degenerate dyon point
 due to the strong $SU(2)$ dynamics.
On the other hand, away from the origin of the moduli space,
 the potential energy decreases as we move toward infinity
 eventually realizing an almost SUSY vacuum.
In addition to this runaway vacuum, there are SUSY vacua on
the Higgs branch which survive quantum corrections. We have
estimated the decay rate of the local minimum
 in the triangle approximation
 and found that the false vacuum is parametrically long-lived.
In this meta-stable vacuum, not only SUSY but also the R-symmetry are
broken.
Interestingly, the basic structure of a messenger sector
 in the gauge mediation scenario is inherent in our model
 in the meta-stable vacuum.
Once the flavor symmetry among massive hypermultiplets
 is gauged as the Standard Model gauge group,
 they play the role of messenger fields
 and supersymmetry breaking is transmitted into the SUSY Standard
 Model sector through the Standard Model gauge interactions.
In order to  obtain a more realistic phenomenological model,
 it is necessary to enlarge the gauge group so as to
 include more flavors.
It is an interesting question whether such a model still
 exhibits a meta-stable vacuum suitable for phenomenology.
This direction is worth investigating in the future. \\\\

\noindent {\large \bf Acknowledgements}

We would like to thank K.~Ohta, S.~Terashima and N.~Yokoi for
their useful comments and in particular for pointing out
the existence of SUSY vacua.
The work of N.~O. is partly supported by
the Grant-in-Aid for Scientific Research in Japan (\#15740164).
S.~S. is supported by the bilateral program of Japan Society
for the Promotion of Science (JSPS) and Academy of Finland, ``Scientist
Exchanges.''

\vspace{10mm}
\begin{appendix}
\section*{Appendix : Derivations of the effective couplings in terms of 
the Weierstrass functions}
\setcounter{equation}{0}
\def\theequation{A.\arabic{equation}}

In this appendix, we exhibit the derivations
 of the effective couplings in term of the Weierstrass functions.
The derivations are applicable for all the case of flavors
($N_f=1,2,3$), so that we shall write the dynamical scale
 $\Lambda_{N_f}$, corresponding to each flavor case.
It is convenient to introduce the uniformization variable $z$
 through the map with the Weierstrass $\wp$ function,
\begin{eqnarray}
 (\wp(z),\wp^\prime(z))=(X,Y)\,.\label{eq:map}
\end{eqnarray}
Using this map, the half period $\omega_i/2$
 is mapped into the root $e_i=\wp(\omega_i/2)$
 ($\omega_3=\omega_1+\omega_2$).
The inverse map is defined as
\begin{eqnarray}
 z_0=\Psi^{-1}(x_0)=\int_{x_0}^\infty\frac{dX}{Y}
 =-\frac{1}{\sqrt{e_2-e_1}}F(\phi,k)\,,
\end{eqnarray}
where we changed the integration variable $X$
 by $t^2=(e_2-e_1)/(X-e_1)$,
 and $F(\phi,k)$ is the incomplete elliptic integral given by
\begin{eqnarray}
F(\phi,k)=\int_0^{\sin\phi}\frac{dt}
          {[(1-t^2)(1-k^2t^2)]^{1/2}} ; \; \;
       \sin^2\phi=\frac{e_2-e_1}{x_0-e_1}\,. \label{incomp}
\end{eqnarray}

We derive the effective couplings, $\tau_{12}$ and $\tau_{11}$,
 by using the map of Eq.~(\ref{eq:map}).
The effective coupling $\tau_{12}$ is described by
\begin{eqnarray}
\tau_{12}=\frac{\partial a_{2D}}{\partial a_1}
          \Bigg{|}_{a_2}
         =\frac{\partial a_{2D}}{\partial a_1}
           \Bigg{|}_u
          -\tau_{22}
          \frac{\partial a_{2}}{\partial a_1}
           \Bigg{|}_u\,. \label{eq:tau12a}
\end{eqnarray}
The partial derivative of the periods $a_{2D}$ and $a_2$
 with respect to $a_1$ can be calculated using
 Eqs.~(\ref{eq:period})-(\ref{eq:f2}) as
\begin{eqnarray}
 \frac{\partial a_{2i}}{\partial a_1}\Bigg{|}_u
  = \oint_{\alpha_i}
       \frac{\partial \lambda_{SW}}{\partial a_1}\Bigg{|}_u
  = Q^{(N_f)}(a_1,~\Lambda_{N_f})
       \int_{e_j}^{e_3}\frac{dX}{2Y(X-c)} \;~~(i\neq j)\,,
\end{eqnarray}
where the coefficient $Q^{(N_f)}$ is given by
\begin{eqnarray}
Q^{(N_f)}(a_1,\Lambda_{N_f})
 =-\frac{N_f(\sqrt{2}a_1)^{N_f-1}\Lambda_{N_f}^{4-N_f}}{16\pi}\,.
\end{eqnarray}
Using the map of Eq.~(\ref{eq:map}), the integral can be described as
\begin{eqnarray}
 \frac{\partial a_{2i}}{\partial a_1}\Bigg{|}_u
  & = &Q^{(N_f)}
       \int_{\omega_j}^{\omega_3}\frac{dz}{2(\wp(z)-\wp(z_0))}
       \nonumber \\
  & = &\frac{Q^{(N_f)}}{2}
       \frac{1}{\wp^\prime(z_0)}
       \left(\log\frac{\sigma(z-z_0)}{\sigma(z+z_0)}
             +2z\zeta(z_0)
       \right)\,,
\end{eqnarray}
where $\wp(z_0)=c$, $\zeta(z)$ is the Weierstrass zeta function,
 and we used the definition of the Weierstrass sigma function,
 $\zeta(z)=\frac{d}{dz}\log\sigma(z)$, and the relation
\begin{eqnarray}
\frac{\wp^\prime(z_0)}{\wp(z)-\wp(z_0)}
        =\zeta(z-z_0)-\zeta(z+z_0)+2\zeta(z_0)\,.
\end{eqnarray}
Taking into account that $Y$ corresponds to $\wp^\prime(z)$
 under the map of Eq.~(\ref{eq:map}),
 the pole $\wp^\prime(z_0)$ can be easily obtained as
\begin{eqnarray}
\wp^\prime(z_0)^2=-\left(\frac{N_f2^{(N_f-1)/2}
                          \Lambda_{N_f}^{4-N_f}}{32}
                   \right)^2\,.
\end{eqnarray}
Using the pseudo periodicity of the Weierstrass sigma function,
\begin{eqnarray}
\sigma(z_0+\omega_i)=
 -\sigma(z_0)\exp\left(2\zeta
  \left(\frac{\omega_i}{2}\right)
       \left(z_0+\frac{1}{2}\omega_i\right)
  \right)\,,
\end{eqnarray}
we obtain
\begin{eqnarray}
\frac{\partial a_{2i}}{\partial a_1}\Bigg{|}_u
 =-\frac{1}{\pi i}
      \left[\omega_i\zeta(z_0)
            -2z_0\zeta\left(\frac{\omega_i}{2}\right)
      \right]\,.\label{eq:anobibun}
\end{eqnarray}
The zeta function at half period can be described by integral
 representations as
\begin{eqnarray}
 \zeta\left(\frac{\omega_i}{2}\right)=-I_2^{(i)}\,.\label{eq:int}
\end{eqnarray}
Substituting Eq.~(\ref{eq:anobibun}) into Eq.~(\ref{eq:tau12a})
 and using the Legendre relation
\begin{eqnarray}
\omega_1\zeta\left(\frac{\omega_2}{2}\right)
 -\omega_2\zeta\left(\frac{\omega_1}{2}\right)
 =i\pi\,,
\end{eqnarray}
we finally obtain
\begin{eqnarray}
\tau_{12}=-\frac{2z_0}{\omega_2}\,.
\end{eqnarray}

Next we derive the effective coupling $\tau_{11}$,
 which is given by differentiating $a_{1D}$ of Eq.~(\ref{eq:mass})
 with respect to $a_1$ with $a_2$ fixed such as
\begin{eqnarray}
\tau_{11} =-
           \int_{x_n^-}^{x_n^+}
           \left[\frac{\partial \lambda_{ SW}} {\partial u}
                 \Bigg{|}_{a_1}
                  \frac{\partial u}
                  {\partial a_1}
                  \Bigg{|}_{a_2}
                +\frac{\partial \lambda_{SW}}
                      {\partial a_1}
                 \Bigg{|}_u
           \right]
           +\frac{\partial \tilde{C}}{\partial a_1}\,. \label{eq:pretau11}
\end{eqnarray}
The integral can be evaluated by using the map (\ref{eq:map}).
Although the integral contains a divergence,
 it can be regularized by using the freedom
 of the integration constant $\tilde{C}$.
Let us demonstrate this regularization
 by introducing the regularization parameter $\epsilon$ as follows.
\begin{eqnarray}
\tau_{11}&=&-\int_{x_0^-+\epsilon}^{x_0^++\epsilon}
             \left[\frac{\partial \lambda_{ SW}} {\partial u}
                 \Bigg{|}_{a_1}
                 \left(-\frac{\partial u}{\partial a_2}\Bigg{|}_{a_1}
                        \frac{\partial a_2}{\partial a_1}\Bigg{|}_{a_2}
                 \right)
                 +\frac{\partial \lambda_{SW}}
                      {\partial a_1}
                 \Bigg{|}_u
             \right]
             +\frac{\partial \tilde{C}}{\partial a_1}
             \nonumber \\
         &=&-\int_{-z_0+\epsilon}^{z_0+\epsilon}dz
             \left[-\frac{1}{\pi i\omega_2}
                 \left(\omega_2\zeta(z_0)
                  -2z_0\zeta\left(\frac{\omega_2}{2}\right)
                 \right)
                 +\frac{Q^{(N_f)}}{4(\wp(z)-\wp(z_0))}
             \right]
             +\frac{\partial \tilde{C}}{\partial a_1}
             \nonumber \\
         &=&-\frac{1}{\pi i}\left(\log\sigma(2z_0)
              -\frac{4z_0^2}{\omega_2}\zeta
               \left(\frac{\omega_2}{2}\right)\right)
              +\frac{1}{\pi}\log\sigma(\epsilon)+\frac{1}{2}
              +\frac{\partial \tilde{C}}{\partial a_1}\,.
\end{eqnarray}
The divergent part, $\log\sigma(\epsilon)$, can be subtracted
 by taking the integration constant such that
 $\tilde{C}=Ca_1-\frac{a_1}{2}-\frac{a_1}{\pi}\log\sigma(\epsilon)$,
 and we finally obtain Eq.~(\ref{eq:tau11})
 with the relation of Eq.~(\ref{eq:int}).
\end{appendix}


\end{document}